\documentclass{ametsocV6.1}

\usepackage{threeparttable}
\usepackage{multirow}
\usepackage{makecell}
\nolinenumbers

\title{Realistic ENSO Dynamics Requires a Damped Nonlinear Recharge Oscillator}

\authors{
Sooman Han,\aff{a}\correspondingauthor{Sooman Han, sooman.han@yale.edu}
Alexey V. Fedorov,\aff{a,b}
and Jérôme Vialard \aff{b}
}

\affiliation{
\aff{a}{Department of Eath and Planetary Sciences, Yale University, New Haven, CT, USA}\\
\aff{b}{LOCEAN-IPSL, Sorbonne Université -CNRS-IRD-MNHN, Paris, France}
}

\abstract{The dynamics of the El Niño–Southern Oscillation (ENSO) are succinctly captured by the Recharge Oscillator (RO) framework. However, to simulate ENSO realistically, careful choices must be made regarding the RO’s key parameters. In particular, nonlinear parameters govern how well the model reproduces ENSO asymmetries—El Niño events tend to be stronger but relatively short, often transitioning into La Niña, whereas La Niña events are typically weaker but may last longer. While amplitude asymmetry has been studied within the RO framework, duration and transition asymmetries remain less explored and their causes are debated. In this study, by systematically exploring the RO parameter space—rather than relying on commonly used fitting methods—we identify optimal parameter values that successfully capture key linear and nonlinear ENSO characteristics. In doing so, we revisit several foundational elements of the RO framework. First, we analytically derive the phase relationship between temperature and heat content anomalies, showing that it depends on the signs of the Bjerknes feedback and the ocean damping timescale. We show that self-sustained oscillations fail to reproduce the observed kurtosis of Niño indices. We further derive an analytical expression for the power spectrum and argue that incorporating red noise forcing, rather than white noise, introduces unnecessary complexity.  The most realistic yet simplest RO configuration is a strongly damped oscillator, with a decay timescale shorter than the dominant period, forced by multiplicative white noise and influenced by weak deterministic nonlinearities. Identifying these minimal components preserves the RO framework’s clarity and isolates the core physical processes underlying ENSO behavior.}

\begin{document}

\maketitle

%
%
%
%
%
%

\statement
The El Niño–Southern Oscillation (ENSO) is a coupled air–sea climate phenomenon in the tropical Pacific with global societal impacts. The Recharge Oscillator (RO) is a simple yet insightful mathematical model that captures key features of ENSO. Nevertheless, standard RO configurations often fail to fully reproduce the observed ENSO behavior. Here, we identify RO configurations that better capture the key linear and nonlinear ENSO characteristics, including the longer duration of La Niña compared to El Niño and the more frequent transition from El Niño to La Niña than the reverse. This study contributes to a more physically consistent and quantitatively realistic RO framework for modeling ENSO.

%

\section{Introduction}\label{sec:intro}

The El Niño-Southern Oscillation (ENSO) is a coupled ocean-atmosphere phenomenon that plays a major role in driving global climate variability on interannual timescales \citep{Trenberth97, Timmermann18}. ENSO events typically occur every 2–7 years and influence global weather patterns, precipitation, and extreme events \citep{Dai00, Cai15, Emerton17}. Consequently, improving our understanding of ENSO dynamics and its predictability is essential for mitigating its substantial socio-economic impacts \citep{Callahan23}.

ENSO is characterized by fluctuations in sea surface temperature (SST) in the central and eastern equatorial Pacific and atmospheric pressure differences at sea level across the Pacific basin, known as the Southern Oscillation \citep{Walker32}. \cite{Bjerknes69} proposed that warm SST anomalies (SSTAs) in the eastern equatorial Pacific weaken the zonal SST gradient, reducing equatorial trade winds. This suppression of trade winds limits upwelling, deepens the eastern equatorial Pacific thermocline, and allows warm surface waters to accumulate, amplifying the initial warming. The weakened trade winds also slow westward surface currents, allowing warm water from the western Pacific to spread eastward, reinforcing SSTA. Enhanced atmospheric convection over the warming region further weakens the Walker Circulation, sustaining reduced trade winds and strengthening the warming signal. Together, these processes contribute to a critical positive feedback, now known as the Bjerknes feedback. Building on this insight, \cite{Wyrtki75, Wyrtki85} highlighted that ENSO’s cyclic nature requires a redistribution of heat content driven by the trade winds. This involves recharge of warm water in the western Pacific before an El Niño event, followed by a discharge of heat after its peak, facilitating the phase transition necessary for ENSO’s periodic behavior \citep{Meinen00}.

Subsequent studies have developed theoretical frameworks of different complexities to explain ENSO dynamics, including the Zebiak-Cane (ZC) model \citep{Zebiak87}, the Delayed Oscillator (DO) framework \citep{Suarez88, Battisti89}, and the Recharge Oscillator (RO) framework \citep{Jin97a} among several others. The ZC model is an intermediate-complexity coupled model that employs a shallow-water reduced-gravity approximation to simulate key ocean-atmosphere interactions. The DO framework  builds on the ZC model, emphasizing the role of positive Bjerknes feedback in amplifying SSTAs and a delayed negative feedback process driven by planetary wave dynamics that eventually terminates El Niño, with SSTAs serving as the primary state variable. The RO framework  extends this understanding by explicitly incorporating the buildup and discharge of heat content in the western Pacific, using SSTA and Ocean Heat Content anomalies (OHCA) as key state variables to describe ENSO's cyclic behavior. Both DO and RO frameworks could be formally derived from the shallow-water equations under the low-frequency approximation \citep{Fedorov10, Stuivenvolt-Allen25a}. Together, these frameworks provide the foundation for understanding ENSO variability \citep{Jin20review}.

The RO has gained significant attention for its ability to quantitatively explain the growth rate and frequency of ENSO using a set of linear parameters that are directly linked to the ocean's mean state and key dynamical processes \citep{Jin06, Kim11a, Kim11b, Kim14, Lu18, Jin20review}. The RO framework exists in various levels of complexity, ranging from the simplest linear version \citep{Burgers05}, to ROs incorporating various nonlinear terms \citep{Jin07, Levine10, An20fokker, Chen20, Kim20, Dommenget23a, Izumo24}, and an extended RO (XRO) that integrates interactions with other ocean basins \citep{Zhao24}. These variations allow the RO framework to capture different aspects of ENSO dynamics, including its amplitude, dominant frequency, seasonal synchronization (ENSO events peak in boreal winter) and positive skewness \citep{Vialard25}. Nevertheless, several ambiguities persist regarding the application of the RO framework to ENSO simulations. This includes the appropriate definition of the OHCA variable, the dynamical regime under which ENSO operates, and the identification of a minimal yet sufficiently effective RO configuration capable of reproducing the observed asymmetries \citep{Vialard25}.

The choice of OHCA variable used in the RO has been questioned by for example \cite{Neske18}. Some studies \citep{Chen20, Jin20review} use the Western Pacific heat content ($h_w$), which was originally used by \cite{Jin97a} based on \cite{Wyrtki75} seminal work, and supported by studies such as \cite{Planton18} and \cite{Izumo19a} for its ability to effectively represent ENSO memory across successive phases. Meanwhile, other studies \citep{Burgers05, Jin07, Levine10, An20fokker, Kim20, Dommenget23a, Zhao24} adopt the equatorial Warm Water Volume (WWV) ($h_{eq}$). \cite{Meinen00} demonstrated that while both $h_{eq}$ and $h_w$ are effective predictors of ENSO, $h_{eq}$ tends to exhibit a higher correlation with the Niño Index at a shorter lead time than $h_w$. Alternatively, some proposed $h_{ind}$ (short for independent) to define a heat content variable that is orthogonal to $T_E$ \citep{Izumo22, Dommenget23b, Priya24}. The first objective of this study is to investigate the consequences of these choices for the RO parameters.

Uncertainties remain regarding the dynamical regime of ENSO—specifically, whether it operates as a self-sustained oscillator \citep{Suarez88, Battisti89, Tziperman94, Tziperman95}, in which a cubic nonlinearity in the SSTA equation limits amplitude growth in an otherwise unstable system, or as a stochastically driven damped (stable) oscillator \citep{Penland96, Moore97, Burgers99a, Thompson01, Fedorov03, Jin07, Zavala-Garay08}, where persistent stochastic forcing is necessary to sustain variability and prevent decay to a quiescent, zero-anomaly state. These two perspectives are commonly referred to as the self-sustained and damped regimes, respectively \citep{Philander03, Fedorov03}. Notably, all regression-based analyses conducted within the RO framework have concluded that ENSO corresponds to a damped system \citep[e.g.,][]{Burgers05, Frauen12, Wengel18}. In contrast, studies that analytically derive growth rates in terms of mean oceanic states and dynamical feedbacks suggest that ENSO may in fact be self-sustained \citep{Kim11a, Kim11b, Kim14}. Most recently, \citet{Weeks25} argued that the regression method itself may introduce biases, rendering it inconclusive for determining the true stability of ENSO. Accordingly, the second objective of this study is to assess whether a damped or self-sustained regime of the RO more accurately reflects ENSO behavior in observations.

Within the RO framework, stochastic noise forcing is typically used to represent high-frequency wind variations, such as Westerly Wind Bursts (WWBs), that are not directly driven by SSTA \citep{Fedorov02, Fedorov03, Lengaigne04, Eisenman05, Fedorov15, Capotondi18, Yu20, Liang21, Yu22}. Previous RO studies have employed either white noise \citep{Burgers05, Frauen12, An20fokker} or red noise with varying decorrelation times, ranging from several days \citep{Vijayeta18, Wengel18, Izumo24} to approximately one month \citep{Jin07, Levine10, Levine15, Levine17, Chen20}. Despite these varying approaches, no study has systematically compared the differences that arise between these noise assumptions. Thus, the third objective of this study is to investigate which noise assumption best aligns with observed ENSO characteristics.

Despite its strengths, the RO framework faces difficulties in capturing ENSO’s asymmetries and nonlinear behaviors \citep{An20review, Jin20review}. Among these, amplitude asymmetry where El Niño events tend to be stronger than La Niña events is the most extensively studied and successfully reproduced. This feature has been attributed to several factors, including the quadratic term in the SSTA equation, which can represent either the SST threshold for deep atmospheric convection, oceanic nonlinearities such as nondynamical heating (NDH) and tropical instability waves (TIWs), or both \citep{An04a, An08, Frauen10, Geng19, Takahashi19, An20fokker, An20review, Jin20review, Kim20, Srinivas24}. It has also been linked to state-dependent atmospheric forcing from WWBs, which tend to be more active during warm phases \citep{Jin07, Levine10}. In contrast, duration asymmetry—where La Niña events typically persist longer and more frequently extend into a second year or beyond, compared to the generally shorter-lived El Niño events \citep{Kessler02, Fedorov03, Okumura10}—and phase transition asymmetry-characterized by the tendency for El Niño events to transition into La Niña but not vice versa \citep{An20review}- remain less well understood. Some studies have proposed additional nonlinear terms that may contribute to duration and phase transition asymmetries \citep{An20fokker, An20review, Izumo24}, but the individual contributions of these terms have not yet been systematically assessed. Therefore, the fourth objective of this study is to identify which forms of nonlinearity within the RO framework are minimally sufficient to reproduce the observed ENSO asymmetries.

Finally, nearly all previous RO studies have relied on parameters obtained through regression fitting, in which temperature and heat content anomalies are treated as independent variables and regressed against their respective tendencies. This method has been applied to estimate both linear and nonlinear parameters. However, as \citet{Weeks25} cautioned, this approach may introduce potential biases. In addition, significant uncertainties remain regarding the appropriate choice of state variables. These include how to define the OHCA variable, as previously discussed, and which SST region (e.g., Niño 3 vs. Niño 3.4) should be used. Furthermore, \citet{Oldenborgh21} and \citet{Izumo19b} suggest that relative SST—defined as SSTA relative to the tropical mean—may offer a more accurate representation of ENSO, introducing further ambiguity in the selection of the temperature variable. In light of these uncertainties, this study deliberately avoids relying on a single parameter set obtained through regression fitting with specific state variables. Instead, we perform a comprehensive parameter sweep to identify the optimal set of model parameters that best reproduces the observed characteristics of ENSO.

Building upon this background, the present study aims to develop a RO “recipe” that retains a minimal set of components, yet effectively captures both the linear and nonlinear characteristics of ENSO by addressing the aforementioned objectives in a systematic, step-by-step manner. In particular, beyond successfully reproducing commonly captured features such as overall amplitude and its asymmetry, dominant frequency, and seasonal synchronization, we specifically aim to reproduce the lagged autocorrelation structure, as well as the duration and transition asymmetries—features that have received comparatively little attention in previous RO studies. Section~\ref{sec:methods} describes the key observed ENSO features to be replicated within the RO framework and outlines the RO equations used in this study. Section~\ref{sec:heat_content} justifies the use of western Pacific heat content for the $h$ variable in our study. Section~\ref{sec:stable_or_unstable} demonstrates that a self-sustained ENSO regime fails to reproduce the observed kurtosis of Niño indices. Section~\ref{sec:white_or_red} shows that using red instead of white noise forcing introduces unnecessary complexity into the model. Section~\ref{sec:nonlinear_parameters} highlights the distinct roles of different nonlinear parameters in generating ENSO asymmetries. Section~\ref{sec:parameter_sweep} systematically explores the parameter space to optimize simulated ENSO properties, successfully reproducing established benchmarks of the RO framework. Finally, Section~\ref{sec:discussion_and_conclusions} discusses the broader implications of these findings and concludes by underscoring the critical importance of atmospheric nonlinearities in shaping ENSO asymmetry.

\section{Methods and Observations}\label{sec:methods}
\subsection{Key ENSO Observations}

\begin{figure}[htbp]
\centering
\includegraphics[width=0.82\textwidth]{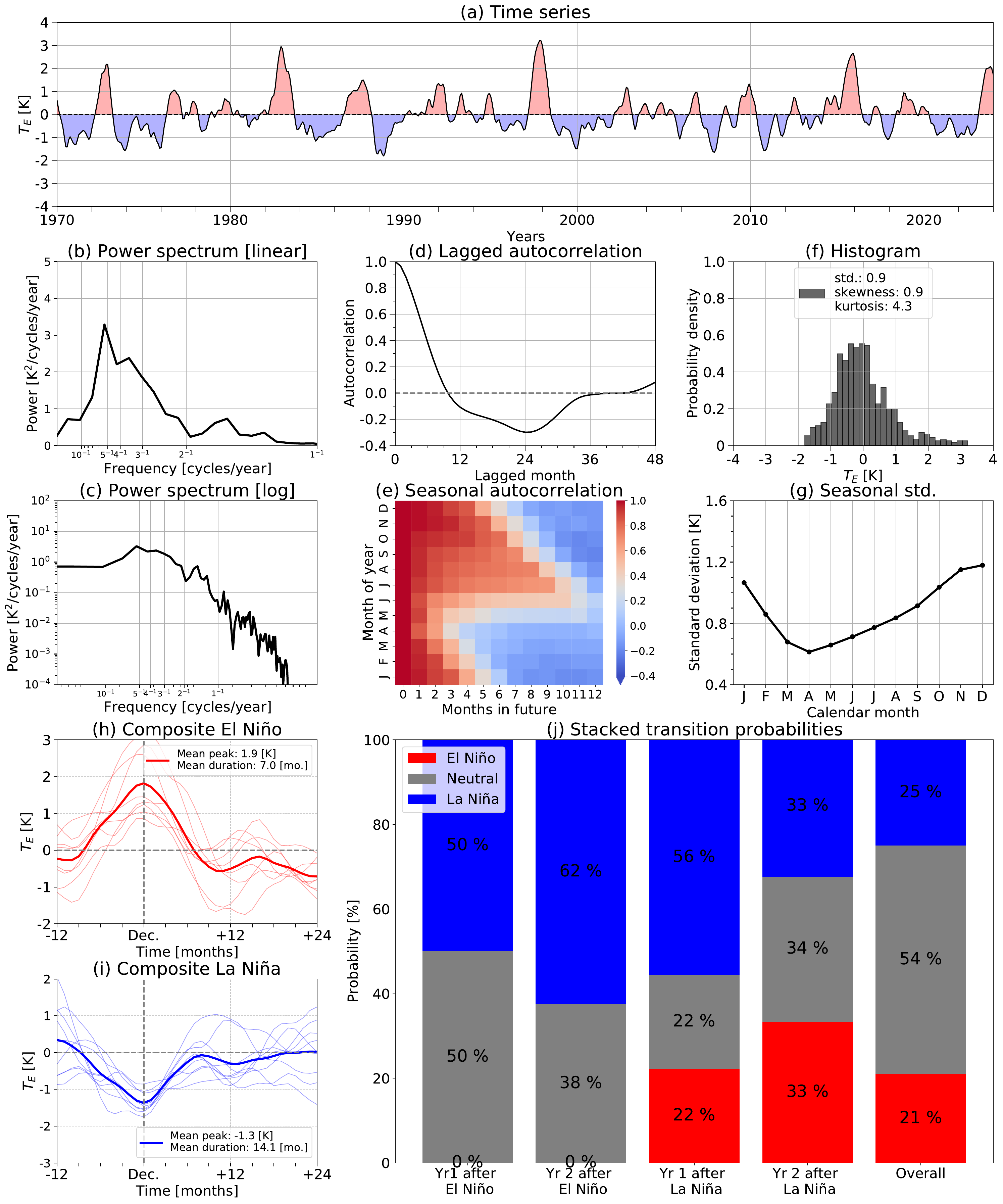}
\caption{Observed ENSO characteristics based on the monthly Niño 3 index (150°W–90°W, 5°S–5°N) from HadISST (1970–2024), smoothed with a 1–2–1 filter as in \cite{Okumura10}: (a) the time series, (b–c) power spectrum in linear and logarithmic scales, (d) lagged autocorrelation, (e) seasonal autocorrelation, (f) histogram, (g) seasonal standard deviations, (h) and (i) composites of moderate-to-strong El Niño and La Niña events defined by the 80th/20th percentiles, and (j) stacked transition probabilities. Transition probabilities in (j) are based on thresholds: El Niño ($T > 0.5$~K), neutral ($|T| \leq 0.5$~K), and La Niña ($T < -0.5$~K) at 1–2 years post-peak. Event duration is defined as the time from peak to when the Niño index drops below 0.1 standard deviations \citep{Choi13}.}
\label{fig:observations}
\end{figure}

Figure~\ref{fig:observations} presents the key observational characteristics of ENSO based on monthly HadISST data spanning the period from 1970 to 2024. We note that the available observations over some 150 years show strong ENSO modulation \citep[e.g.,][]{Fedorov20review}, but the last 50 or so years of the record represent the most reliable data. Figure~\ref{fig:observations}(a) shows the time series of the Niño 3 index, defined as SSTA averaged over the region 150°W–90°W and 5°S–5°N. ENSO exhibits pronounced asymmetry: El Niño events are typically stronger and shorter-lived, while La Niña events tend to have smaller amplitudes but persist for multiple years \citep{Okumura10, An20review}. Figure~\ref{fig:observations}(b) and (c) display the power spectral density of ENSO in linear and logarithmic scale, showing a broad interannual spectral peak \citep{Rasmusson82}. Figure~\ref{fig:observations}(d) shows the lagged autocorrelation of the Niño 3 index, which has been interpreted as evidence that ENSO behaves like a damped oscillator \citep{Burgers99a}. Figure~\ref{fig:observations}(e) illustrates the month-by-month autocorrelation structure of ENSO, which exhibits a sharp decline during boreal spring. This reduction in persistence reflects the decreased memory of initial anomalies, a phenomenon known as the spring predictability barrier \citep{Lopez14, Levine15}. Figure~\ref{fig:observations}(f) displays the data histogram, which encapsulates both amplitude and duration/transition asymmetries collectively through its positive skewness \citep{Burgers99b}. Figure~\ref{fig:observations}(g) presents the seasonality of ENSO standard deviations, revealing that ENSO is phase-locked to peak during boreal winter in the Northern Hemisphere \citep{Rasmusson82}.

Figures~\ref{fig:observations}(h) and \ref{fig:observations}(i) present the composite evolution of moderate-to-strong El Niño and La Niña events, defined as Niño index values exceeding the 80th percentile for El Niño and falling below the 20th percentile for La Niña—corresponding to thresholds of approximately 0.9~K and –0.9~K, respectively. This thresholding approach is similar to that used by \cite{Choi13} and \cite{Chen20}, focusing on moderate-to-strong events to better highlight the pronounced nonlinearities and asymmetries they tend to exhibit. Event duration is defined as the time from its peak to termination, when the Niño index falls below 25\% of the standard deviation from the annual mean, following \citet{Choi13}. The asymmetric evolution is evident, with El Niño events exhibiting greater amplitude and shorter duration (1.9~K and 7.0~months) compared to La Niña events (–1.3~K and 14.1~months).

Figure~\ref{fig:observations}(j) presents stacked transition probabilities, illustrating the likelihood of transitioning to El Niño ($T > 0.5$~K), La Niña ($T < -0.5$~K), or neutral conditions ($|T| \leq 0.5)$~K) one and two years after a moderate-to-strong ENSO event. Note that the definitions of El Niño, La Niña, and neutral conditions used in the stacked transition probability plots are not limited to moderate-to-strong events. Transition probabilities are calculated using the conventional classification in order to avoid biasing the statistics toward neutral conditions. This inclusive approach is essential for accurately interpreting the prevalence of El Niño or La Niña phases in the years following each moderate-to-strong ENSO event. It reveals that La Niña is the dominant state both one and two years following moderate-to-strong El Niño events, accounting for 50\% and 62\% of cases— in stark contrast to the 0\% probability for El Niño (first and second bars in Figure~\ref{fig:observations}(j)). 56\% of moderate-to-strong La Niña events persist into a second consecutive year (the third bar), indicating a strong tendency for multi-year La Niñas \citep{Okumura10}. Around 33\% evolve into a third-year ("triple dip") La Niña (the fourth bar), such as in 2020-2023. For comparison, the overall probabilities of El Niño, neutral, and La Niña conditions during the boreal winter are shown based on the observational record (the fifth bar).

These features underscore ENSO’s distinct linear as well as nonlinear characteristics—namely, amplitude, duration, and phase transition asymmetries. This study aims to reproduce these behaviors within the RO framework by systematically investigating the roles of nonlinear terms and optimizing both the equation structure and parameter values.

\subsection{The Recharge Oscillator Equations}

We consider the following form of the nonlinear RO (c.f. \cite{Jin20review, Vialard25}):

\begin{linenomath*}
\begin{equation}
\frac{dT_E}{dt} = RT_E + F_1 h + b_T T_E^2 - c_T T_E^3 + d_T T_E h + \sigma_T \xi_T (1 + BH(T_E) T_E) 
\end{equation}

\begin{equation}
\frac{dh}{dt} = -\varepsilon h - F_2 T_E - b_h T_E^2 + \sigma_h \xi_h
\end{equation}
\end{linenomath*}

\noindent The variables $T_E$ and $h$ represent SSTA in the eastern Pacific and an equatorial OHCA, still to be appropriately defined, with units of K and m, respectively. For $T_E$, we will consider the Niño 3 index, as shown in Figure~\ref{fig:observations}, rather than the Niño 3.4 index, since the Niño 3 index exhibits higher skewness, making ENSO's asymmetric features more clearly distinguishable \citep{An04a, Hayashi17}. The best choice of OHCA as a state variable will be discussed in the following section. The linear parameters—$R$, $F_1$, $\varepsilon$, and $F_2$—correspond to the Bjerknes feedback, delayed oceanic feedback efficiency, ocean adjustment rate, and recharge/discharge efficiency, with units of month$^{-1}$, K$\cdot$m$^{-1}\cdot$month$^{-1}$, month$^{-1}$, and m$\cdot$K$^{-1}\cdot$month$^{-1}$, respectively. When $R$ includes seasonal modulation, it is expressed as 

\begin{linenomath*}
\begin{equation}
R = R_0 - R_a \cos(\omega_a t - \phi) 
\end{equation}
\end{linenomath*}

\noindent where $R_0$ and $R_a$ are the steady and seasonal components, $\omega_a= 2\pi/12$ is the annual frequency in months$^{-1}$, and $\phi$ denotes the phase lag between the seasonal modulation and SSTA growth \citep{Kim21}. In this study, we consider only the steady component in $R$ in Sections~\ref{sec:heat_content} through \ref{sec:nonlinear_parameters}, while the seasonal component in $R$ is introduced in Section~\ref{sec:parameter_sweep} to simulate realistic ENSO features such as seasonal phase locking.

The deterministic nonlinear parameters $b_T$, $c_T$, $d_T$, and $b_h$ have units of K$^{-1}\cdot$month$^{-1}$, K$^{-2}\cdot$month$^{-1}$, m$^{-1}\cdot$month$^{-1}$, and m$\cdot$K$^{-2}\cdot$month$^{-1}$, respectively. Specifically, $b_T$ represents asymmetry in the atmospheric convective response to warm versus cold SST anomalies (an atmospheric nonlinearity), and also captures oceanic nonlinearities such as NDH, TIWs, and thermocline feedback \citep{An04a, An08, Frauen10, Geng19, Izumo19b, Takahashi19, An20fokker, An20review, Jin20review, Kim20, Srinivas24, Vialard25}. $c_T$ is a cubic nonlinearity, physically motivated by a Taylor expansion of thermocline depth anomalies \citep{Battisti89, An08}, and prevents divergence of the system as a nonlinear damping. The parameter $d_T$ similarly reflects oceanic nonlinear feedbacks similar to $b_T$ \citep{An20fokker, An20review, Kim20}. Lastly, $b_h$ enhances the discharge of heat content following El Niño events, playing a key role in generating La Niña persistence \citep{Izumo24}.

The stochastic terms $\xi_T$ and $\xi_h$ are noise time series with unit variance \citep{Burgers99a, Jin20review, Vialard25}. For white noise, $\xi_T = \dot{W_T}$ and $\xi_h = \dot{W_h}$ are uncorrelated Gaussian white noise processes with units of month$^{-0.5}$ where $W_T$ and $W_h$ denote Wiener process. Hereafter, we use $w_T$ and $w_h$ to denote $\dot{W}_T$ and $\dot{W}_h$, respectively. The noise amplitudes $\sigma_T$ and $\sigma_h$ for white noise have units of K$\cdot$month$^{-0.5}$ and m$\cdot$month$^{-0.5}$, respectively. When modeled as red noise, they satisfy the stochastic differential equations:

\begin{linenomath*}
\begin{equation}
\frac{d\xi_T}{dt} = -m_T \xi_T + \sqrt{2m_T} \, w_T
\end{equation}

\begin{equation}
\frac{d\xi_h}{dt} = -m_h \xi_h + \sqrt{2m_h} \, w_h
\end{equation}
\end{linenomath*}

\noindent where $m_T$ and $m_h$ represent the decorrelation rates with units of month$^{-1}$. In this formulation, the decorrelation time parameter $m_{i=T,h}$ acts as a damping factor for the noise time series. To ensure that the noise term $\xi_{i=T,h}$ maintains unit variance, a scaling factor of $\sqrt{2m_{i=T,h}}$ is introduced in the red noise expression. This scaling ensures that the variance is calculated as: $\frac{(\sqrt{2m_{i=T,h}})^{2}}{2 \times m_{i=T,h}}=1$. Due to the presence of $m_{i=T,h}$, $\xi_T$ and $\xi_h$ are rendered unitless, therefore the amplitudes $\sigma_T$ and $\sigma_h$ take on units of K$\cdot$month$^{-1}$ and m$\cdot$month$^{-1}$, respectively.

The multiplicative noise parameter $B$, which captures the modulation of WWBs by ENSO, has units of K$^{-1}$ \citep{Levine10, Jin20review, Vialard25}. The Heaviside function $H(T_E)$ is defined as:

\begin{linenomath*}
\begin{equation}
H(T_E) =
\begin{cases}
1 & \text{if} \ T_E > 0, \\
0 & \text{if} \ T_E \leq 0,
\end{cases}
\end{equation}
\end{linenomath*}

\noindent $H(T_E)$ is dimensionless, with no associated units.

The parameters in the RO equations have differing physical units. To enable meaningful comparison and interpretation of their relative magnitudes, we normalize the system using dimensionless variables defined by $T_E = T_E / \text{std}(T_E)$ and $h = h / \text{std}(h)$, where $\text{std}(\cdot)$ denotes the standard deviation. This normalization ensures that all linear and nonlinear parameters are expressed in common units of month$^{-1}$ (see Appendix~A for details), and facilitates consistent stepwise changes in parameter values during the parameter sweep simulations. All subsequent simulations are performed using normalized units, with results converted back to original units for analysis. The stochastic RO equations are solved numerically using the Euler–Maruyama scheme with a time step of 0.01 months, and results are recorded at an output interval of 1 month.

\section{Choosing the Ocean Heat Content Variable}
\label{sec:heat_content}  

As a first step in formulating the optimal RO configuration, we explore the simplest form of the RO model—namely, the linear version—and focus our discussion on the OHCA variable. Here, Eqs.~(1) and (2) retain only linear terms: $\frac{dT_E}{dt} = R T_E + F_1 h $ and $\frac{dh}{dt} = -\varepsilon h - F_2 T_E$ \citep{Burgers05}. Table~1 lists the linear parameter values from previous studies. When dimensional values are used, $F_1$ and $F_2$ differ not only in units—K$\cdot$m$^{-1}\cdot$month$^{-1}$ for $F_1$ and m$\cdot$K$^{-1}\cdot$month$^{-1}$ for $F_2$—but also in magnitude, with $F_2$ typically exceeding $F_1$ by a factor of approximately 100. To enable meaningful comparison of parameter sets across studies, we present normalized values. For studies that report only dimensional parameters, linear terms are scaled using $\mathrm{std}(T_E) = 0.88$ and $\mathrm{std}(h_w) = 8.0$, consistent with our analysis. This approach is justified given that $T_E$—whether defined as Niño 3 or Niño 3.4—has comparable magnitude across studies, and as noted by \citet{Dommenget23a, Dommenget23b}, the standard deviations of $h_{eq}$ and $h_{ind}$ are similar to that of $h_w$. Original (unnormalized) parameter values are shown in parentheses.

\begin{figure}[htbp]
  \centering
  \includegraphics[width=1.0\textwidth]{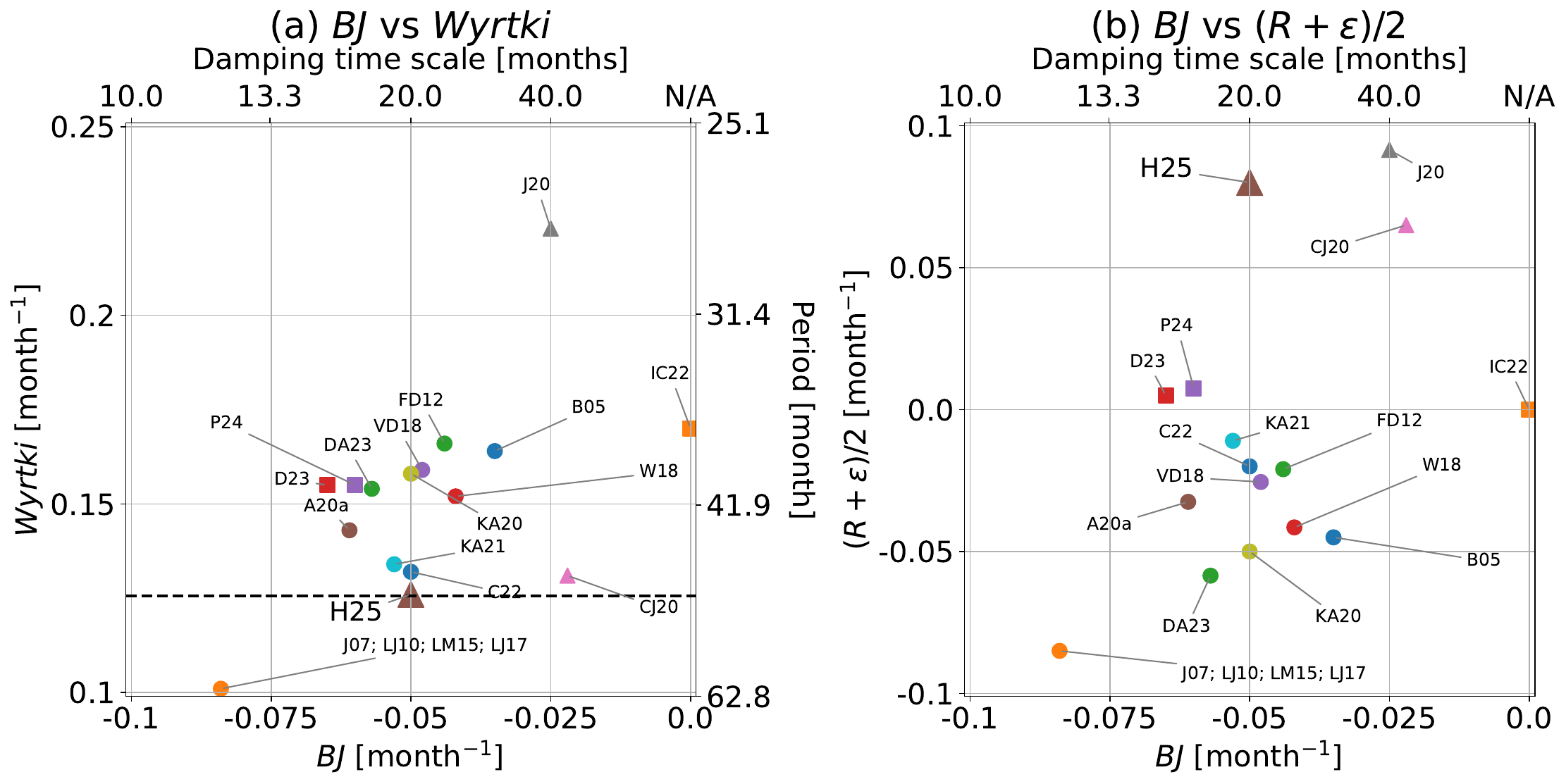}
  \caption{Comparison of key ENSO indices from previous studies and from this study within the RO framework corresponding to Table~1: 
  Panel (a) presents $BJ = \frac{R - \varepsilon}{2}$ vs. $Wyrtki = \sqrt{F_1 F_2 - \frac{(R + \varepsilon)^2}{4}}$. The horizontal dashed line indicates the observed dominant period of 50 months ($Wyrtki=0.126$ month$^{-1}$), estimated from the minimum of the lagged autocorrelation shown in Figure~\ref{fig:observations}(d) (cf. \cite{Jiang21}). Panel (b) shows $BJ = \frac{R - \varepsilon}{2}$ vs. $\frac{R + \varepsilon}{2}$. As illustrated in Figure~\ref{fig:phase_diagrams}, $R + \varepsilon$ controls the sign of the phase space diagram slope and is thus linked to the choice of $h$: $R + \varepsilon > 0 \ \Leftrightarrow \ h = h_w$ (triangle), $R + \varepsilon < 0 \ \Leftrightarrow \ h = h_{eq}$ (circle), and $R + \varepsilon \approx 0 \ \Leftrightarrow \ h = h_{\text{ind}}$ (square). The division by 2 in $\frac{R + \varepsilon}{2}$ is applied to maintain a comparable scale with the $BJ$ index. ENSO indices derived from the optimal parameters from this study are highlighted with larger symbols (brown triangles).
  }
  \label{fig:previous_comparison}
\end{figure}

The $BJ$ and $Wyrtki$ indices, defined as $(R - \varepsilon)/2$ and $\sqrt{F_1 F_2 - (R + \varepsilon)^2/4}$, respectively, describe ENSO growth rates and dominant frequency \citep{Jin06, Lu18}. The expressions for these indices remain unchanged regardless of whether normalized or dimensional variables are used (see Appendix~A for details). In this study, following previous literature \citep[e.g.,][]{Fedorov03, Philander03, Vialard25}, the damped regime is defined in a linear sense, that is, the regime in which the annual-mean component of the BJ index is negative, i.e., $(R_0 - \varepsilon)/2 < 0$, regardless of whether the cubic nonlinearity $c_T$ is present. Although $c_T$ is not required in this regime, its inclusion help eliminate potential runaway effects that can be occasionally produced by a combination of noise and other nonlinear terms. Conversely, the self-sustained regime is defined by $(R_0 - \varepsilon)/2 > 0$. In this regime, the presence of $c_T$ is a necessary condition to ensure that the system remains bounded. Figure~\ref{fig:previous_comparison}(a) illustrates the $BJ$ and $Wyrtki$ indices corresponding to the linear parameters in previous studies. The $BJ$ index generally ranges between $-0.075$ and $-0.025$ month$^{-1}$, consistently indicating a stable or damped ENSO regime. The $Wyrtki$ index typically falls between 0.13 and 0.17 month$^{-1}$, corresponding to periodicities of approximately 36 to 48 months.

Table~1 shows that the signs of the parameters $R$ and $\varepsilon$ vary across studies; in some cases, they are positive, while in others, negative. The key role of the Bjerknes feedback in ENSO suggests that $R$ may be positive at least during some phases of the seasonal cycle, but imposes no constraint on the annual-average value, whose sign may vary depending on how the subprocesses contributing to the Bjerknes feedback are represented \citep{Kim11a, Kim14, Jin20review}. A positive $\varepsilon$ is consistent with dissipative processes, such as energy loss associated with poleward-propagating coastal Kelvin waves at the eastern boundary \citep{Jin97b, Izumo19a}. However, a few studies report slightly negative values of $\varepsilon$, which is physically problematic, as it would imply self-amplifying ocean wave dynamics.

In Table~1, we also listed the choice of $h$ reported in each study and the corresponding sign of $R + \varepsilon$, revealing notable relationships. Studies that adopted $h_w$ report $R + \varepsilon > 0$ \citep{Chen20, Jin20review}. In contrast, studies that employed $h_{eq}$ consistently report $R + \varepsilon < 0$ \citep{Burgers05, Jin07, Levine10, Frauen12, Levine15, Levine17, Wengel18, Vijayeta18, An20fokker, Kim20, Kim21, Crespo22, Dommenget23a}. Finally, studies that adopted $h_{ind}$ show $R+\varepsilon \approx 0$ \citep{Izumo22, Dommenget23b, Priya24}. Figure~\ref{fig:previous_comparison}(b) visually illustrates the sign of $R + \varepsilon$ reported in previous studies.

\begin{table}[htbp]
\begin{center}
\begin{threeparttable}
\caption{Summary of linear parameter values used in RO models, the corresponding choices of heat content variable, and the sign of $R + \varepsilon$ across selected studies. For consistent comparison, all parameters are presented in normalized units month$^{-1}$. When standard deviations of $T_E$ and $h$ were not reported, we adopt $\mathrm{std}(T_E) = 0.88$ and $\mathrm{std}(h_w) = 8.0$ for normalization. These values are assumed to be representative of cases using $h_{eq}$ and $h_{ind}$ as the heat content variable \citep[e.g.,][]{Dommenget23a, Dommenget23b}. See Appendix~A for details. Values in parentheses indicate the unnormalized parameters.}
\begin{tabular}{lcccccc}  
\hline\hline
Reference  & $R$ & $\varepsilon$ & $F_1$ & $F_2$ & $h$ & $R+\varepsilon$ \\

\hline
\hline
\cite{Burgers05} (B05) & -0.08 & -0.01 & 0.17 & 0.17 & $h_{eq}$ & $<0$ \\  \hline

\multirow{3}{*}{\makecell[l]{\cite{Jin07} (J07) \\ \cite{Levine10} (LJ10) \\ \cite{Levine15} (LM15)\\ \cite{Levine17} (LJ17)}} 
& \multirow{3}{*}{-0.17} 
& \multirow{3}{*}{0.00} 
& \multirow{3}{*}{0.13} 
& \multirow{3}{*}{0.13} 
& \multirow{3}{*}{\makecell[l]{$h_{eq}$}} 
& \multirow{3}{*}{$<0$} \\ \\ \\ \hline

\multirow{2}{*}{\makecell[l]{\cite{Frauen12} (FD12)}} 
& \multirow{2}{*}{-0.07} 
& \multirow{2}{*}{0.02} 
& \multirow{2}{*}{\makecell[l]{~0.27\\(0.03)}}  
& \multirow{2}{*}{\makecell[l]{~0.12\\(1.13)}} 
& \multirow{2}{*}{\makecell[l]{$h_{eq}$}} 
& \multirow{2}{*}{$<0$} \\ \\ \hline

\multirow{2}{*}{\makecell[l]{\cite{Wengel18} (W18)}} 
& \multirow{2}{*}{-0.08} 
& \multirow{2}{*}{0.00} 
& \multirow{2}{*}{\makecell[l]{~0.18\\(0.02)}}  
& \multirow{2}{*}{\makecell[l]{~0.16\\(1.46)}} 
& \multirow{2}{*}{\makecell[l]{$h_{eq}$}} 
& \multirow{2}{*}{$<0$} \\ \\ \hline

\multirow{2}{*}{\makecell[l]{\cite{Vijayeta18} (VD18)}} 
& \multirow{2}{*}{-0.07} 
& \multirow{2}{*}{0.02} 
& \multirow{2}{*}{\makecell[l]{~0.18\\(0.02)}}  
& \multirow{2}{*}{\makecell[l]{~0.14\\(1.23)}} 
& \multirow{2}{*}{\makecell[l]{$h_{eq}$}} 
& \multirow{2}{*}{$<0$} \\ \\ \hline

\multirow{2}{*}{\makecell[l]{\cite{An20fokker} (A20a)}} 
& \multirow{2}{*}{-0.09} 
& \multirow{2}{*}{0.03} 
& \multirow{2}{*}{\makecell[l]{~0.18\\(0.02)}}  
& \multirow{2}{*}{\makecell[l]{~0.11\\(1.02)}} 
& \multirow{2}{*}{\makecell[l]{$h_{eq}$}} 
& \multirow{2}{*}{$<0$} \\ \\ \hline

\cite{Chen20} (CJ20) & 0.04 & 0.09 & 0.15 & 0.15 & $h_{w}$ & $>0$ \\ \hline

\multirow{2}{*}{\makecell[l]{\cite{Jin20review} (J20)}} 
& \multirow{2}{*}{0.07} 
& \multirow{2}{*}{0.12} 
& \multirow{2}{*}{\makecell[l]{~0.36\\(0.04)}}  
& \multirow{2}{*}{\makecell[l]{~0.17\\(1.52)}} 
& \multirow{2}{*}{\makecell[l]{$h_{w}$}} 
& \multirow{2}{*}{$>0$} \\ \\ \hline

\multirow{2}{*}{\makecell[l]{\cite{Kim20} (KA20)}} 
& \multirow{2}{*}{-0.10} 
& \multirow{2}{*}{0.00} 
& \multirow{2}{*}{\makecell[l]{~0.18\\(0.02)}}  
& \multirow{2}{*}{\makecell[l]{~0.13\\(1.15)}} 
& \multirow{2}{*}{\makecell[l]{$h_{eq}$}} 
& \multirow{2}{*}{$<0$} \\ \\ \hline

\multirow{2}{*}{\makecell[l]{\cite{Kim21} (KA21)}} 
& \multirow{2}{*}{-0.07} 
& \multirow{2}{*}{0.03} 
& \multirow{2}{*}{\makecell[l]{~0.18\\(0.02)}}  
& \multirow{2}{*}{\makecell[l]{~0.10\\(0.91)}} 
& \multirow{2}{*}{\makecell[l]{$h_{eq}$}} 
& \multirow{2}{*}{$<0$} \\ \\ \hline

\multirow{2}{*}{\makecell[l]{\cite{Crespo22} (C22)}} 
& \multirow{2}{*}{-0.07} 
& \multirow{2}{*}{0.03} 
& \multirow{2}{*}{\makecell[l]{~0.18\\(0.02)}}  
& \multirow{2}{*}{\makecell[l]{~0.12\\(1.05)}} 
& \multirow{2}{*}{\makecell[l]{$h_{eq}$}} 
& \multirow{2}{*}{$<0$} \\ \\ \hline

\cite{Izumo22}\footnotemark[1] (IC22) & 0.00 & 0.00 & 0.17 & 0.17 & $h_{ind}$ & $\approx0$ \\
\hline
\cite{Dommenget23a} (DA23) & -0.12 & -0.00 & 0.17 & 0.16 & $h_{eq}$ & $<0$ \\
\hline
\cite{Dommenget23b}\footnotemark[2] (D23) & -0.06 & 0.07 & 0.15 & 0.16 & $h_{ind}$ & $\approx0$ \\
\hline
\cite{Priya24}\footnotemark[2] (P24) & -0.05 & 0.07 & 0.15 & 0.16 & $h_{ind}$ & $\approx0$ \\
\hline
Han et al. (2025) (H25; this study) & 0.03 & 0.13 & 0.14 & 0.16 & $h_{w}$ & $>0$ \\
\hline
\hline
\end{tabular}
\begin{tablenotes}
\item[a] Uses a combination of equatorial and southwestern WWV.
\item[b] Uses the maximum temperature gradient.
\end{tablenotes}
\end{threeparttable}
\end{center}
\label{tab:previous_results}
\end{table}

\begin{figure}[htbp]
  \centering
  \includegraphics[width=1.0\textwidth]{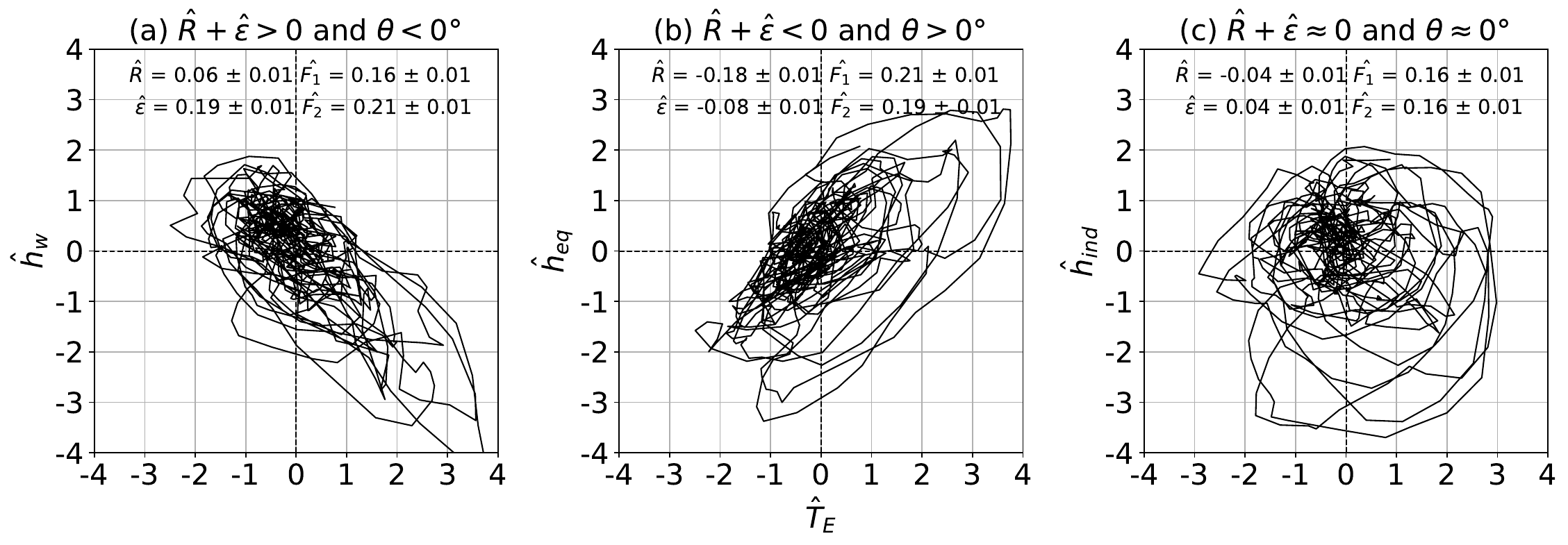}
  \caption{
    $T_E$ and $h$ phase space diagrams for different choices of $h$ (OHCA): (a) $h_w$, representing Western Pacific ocean heat content (120°E–180°E; 5°S–5°N); (b) $h_{eq}$, indicating equatorial WWV (120°E–280°E; 5°S–5°N); and (c) $h_{\text{ind}}$, defined as a mixture of equatorial and southwestern WWV that is orthogonal to $T_E$ (120°E–280°E; 5°S–5°N + 120°E–190°E; 15°S–5°S), after \cite{Izumo22}. The diagrams use monthly HadISST and ORAS5 data for the past 50 years (1970-2020) \citep{Rayner03, Zuo19}. Linear RO parameter values are obtained under linear assumptions for each case. These panels illustrate how the slope sign, $\theta=\frac{1}{2}\arctan\left(\frac{(-\hat{R}+\hat{\varepsilon})}{\hat{F_1}-\hat{F_2}}\right)$, varies depending on the choice of $h$ and its relationship with the sign of $\hat{R} + \hat{\varepsilon}$. Both $T_E$ and $h$ are normalized by their respective standard deviations to ensure comparable scales, as indicated by the hat notation. See Appendices A and B for the normalization procedures and the analytical derivation of the slope expression in terms of linear parameters.
  }
  \label{fig:phase_diagrams}
\end{figure}

This trend is further illustrated in Figure~\ref{fig:phase_diagrams}, which compares $T_E$–$h$ phase space diagrams constructed using $T_E$ from HadISST data and $h_w$, $h_{eq}$, and $h_{ind}$ from ORAS5 data in our study. The corresponding linear parameter values were estimated by performing regression fits using $T_E$ and $h$ as independent variables, regressed against their respective tendencies. The phase space diagram behavior — where $h_w$ is linked with $R + \varepsilon > 0$, $h_{eq}$ is linked with $R + \varepsilon < 0$, and $h_{ind}$ is linked with $R + \varepsilon \approx 0$ — aligns with the parameter trends presented in Table~1, reinforcing the connection between the choice of $h$ and the sign of $R + \varepsilon$.

To understand why the relationship between the choice of $h$ and $R + \varepsilon$ arises, we analytically derived an expression for the slope in the phase space diagram (Appendix~B):

\begin{linenomath*}
\begin{equation}
\theta = \frac{1}{2} \arctan\left(\frac{-(\hat{R} + \hat{\varepsilon})}{\hat{F_1} - \hat{F_2}} \right)
\end{equation}
\end{linenomath*}

\noindent where hat notations indicate normalized parameters. It is important to note that both the numerator and denominator contain information about the angular quadrant, and preserving their respective signs is essential for correctly determining the phase. For the tangent function, the sign of $\theta$ depends only on the sign of the numerator, i.e., $-(\hat{R} + \hat{\varepsilon})$. This implies:

\begin{linenomath*}
\begin{equation}
h=h_w \iff \theta < 0 \iff \hat{R} + \hat{\varepsilon} > 0  
\end{equation}

\begin{equation}
h=h_{eq} \iff \theta > 0 \iff \hat{R} + \hat{\varepsilon} < 0  
\end{equation}

\begin{equation}
h=h_{ind} \iff \theta \approx 0 \iff \hat{R} + \hat{\varepsilon} \approx 0  
\end{equation}

\end{linenomath*}

This means that not only does $R + \varepsilon$ ($\hat{R} + \hat{\varepsilon}$) influence the dominant ENSO period through the Wyrtki frequency (Eq.~A18), but it also governs the phase relationship between the two state variables in the RO equations. This also suggests that using $h_w$ as the heat content variable is a necessary (but not sufficient) condition for ensuring that both $R$ and $\varepsilon$ remain positive. Indeed, adopting $h_{eq}$ as the heat content variable leads to at least one of $R$ or $\varepsilon$ being negative, while using $h_{ind}$ results in $R$ and $\varepsilon$ having opposing signs with similar magnitudes (See Table~1 and Figure~\ref{fig:phase_diagrams}). In other words, using $h_w$ as the heat content variable appears to be the safest way to ensure a positive value of $\varepsilon$, representing an OHCA that decays in the absence of external forcing. We emphasize that this reasoning does not imply any inherent superiority of using $h_w$ over $h_{eq}$ or $h_{ind}$. Rather, adopting $h_w$ as the heat content variable serves to constrain the parameter space to $R > 0$ and $\varepsilon > 0$, thereby ensuring a more physically consistent formulation.

Overall, despite differences in the choice of $h$ and variations in parameter prescriptions across studies—including those using only linear terms \citep{Burgers05, Frauen12, Wengel18, Vijayeta18, Crespo22, Izumo22, Priya24}, those incorporating nonlinearities \citep{Jin07, Levine10, Levine15, Levine17, An20fokker, Chen20, Kim20, Kim21, Dommenget23a, Dommenget23b}, and those including seasonality in $R$ \citep{Levine15, Chen20, Kim21}—there is notable consistency in the resulting $BJ$ and $Wyrtki$ indices (Figure~\ref{fig:previous_comparison}(a)), indicating the robustness of these metrics.

\section{Is ENSO Damped or Self-sustained?}
\label{sec:stable_or_unstable}

In this section, we examine RO simulations under both damped and self-sustained assumptions. While regression-based studies consistently suggest that ENSO behaves as a stable, damped system (with negative $BJ$ values; see Figure~\ref{fig:previous_comparison}), several GCM-based analyses, which compute the $BJ$ index from mean-state variables, suggest that the index can be positive \citep{Kim11a, Kim11b, Kim14, Heede23b}, indicating a self-sustained system, even though it is unclear whether these estimates are reliable \citep{Weeks25}. Here, we explore alternative methods to assess whether ENSO is better characterized by a damped or self-sustained internal model in the system.

Table~\ref{tab:unstable1} outlines the simulation setups. For the Damped case, linear parameters are selected to closely match those from the regression results in Figure~\ref{fig:phase_diagrams}(a), yet simplified for clarity while ensuring the Wyrtki frequency corresponds to a 48-month dominant periodicity, as indicated by the observed autocorrelation structure (Figure~\ref{fig:observations}(d) following \cite{Jiang21}). Additionally, $R$ and $\varepsilon$ are chosen such that the resulting BJ index, $(R - \varepsilon)/2$, corresponds to a typical value reported in previous studies, $-0.05$~month$^{-1}$ (see Figure~\ref{fig:previous_comparison}(a)). In the Self-sustained 1 case, the values of $R$ and $\varepsilon$ are swapped to induce instability, while keeping $R + \varepsilon$ constant in order to preserve the same Wyrtki index (see Eq.~A18). Additionally, a cubic nonlinearity is included to prevent divergence of the system. In the Self-sustained 2 case, we consider a more realistic self-sustained configuration that includes $b_T$, which is capable of inducing ENSO asymmetry. This setup is based on the self-sustained case presented in Table~S1 of \cite{Weeks25}, with the modification of using white noise instead of red noise forcing.

\begin{table}[htbp]
\caption{Equation system and model parameters used to compare damped and self-sustained regimes in RO simulations. Parameter values are given in normalized units. The Self-sustained 2 case is based on parameters from \citet{Weeks25}, normalized and modified to use white noise forcing instead of red noise. Values in parentheses indicate the unnormalized parameters.}
\begin{center}
\begin{tabular}{lcccccccc}

\hline \hline

\multicolumn{9}{c}{Equations} \\ 
\hline
\\
\multicolumn{9}{l}{\makecell[c]{
$\frac{dT_E}{dt} = RT_E + F_1 h +b_T T_E^2- c_T T_E^3 + \sigma_T w_T$ \\ \\
$\frac{dh}{dt} = -\varepsilon h - F_2 T_E + \sigma_h w_h$
}} 
\\ \\ \hline \hline

 & $R$ & $\varepsilon$ & $F_1$ & $F_2$ & $b_T$ & $c_T$ & $\sigma_T$ & $\sigma_h$
\\ \hline

Damped
& 0.10
& 0.20
& 0.20
& 0.20
& 0.00
& 0.00
& 0.20
& 0.20

\\ \hline

Self-sustained 1 
& 0.20
& 0.10
& 0.20
& 0.20
& 0.00
& 0.02
& 0.20
& 0.20



\\ \hline

\multirow{2}{*}{Self-sustained 2}  
& \multirow{2}{*}{0.038} 
& \multirow{2}{*}{0.022}  
& \multirow{2}{*}{\makecell[l]{~0.155 \\ (0.017)}} 
& \multirow{2}{*}{\makecell[l]{~0.138 \\ (1.250)}} 
& \multirow{2}{*}{\makecell[l]{~0.020 \\ (0.058)}} 
& \multirow{2}{*}{\makecell[l]{~0.088 \\ (0.113)}}
& \multirow{2}{*}{\makecell[l]{~0.193 \\ (0.219)}} 
& \multirow{2}{*}{\makecell[l]{~0.144 \\ (1.150)}} 

\\ \\ \hline \hline

\end{tabular}
\end{center}
\label{tab:unstable1}
\end{table}

\begin{figure}[htbp]
  \centering
  \includegraphics[width=1.0\textwidth]{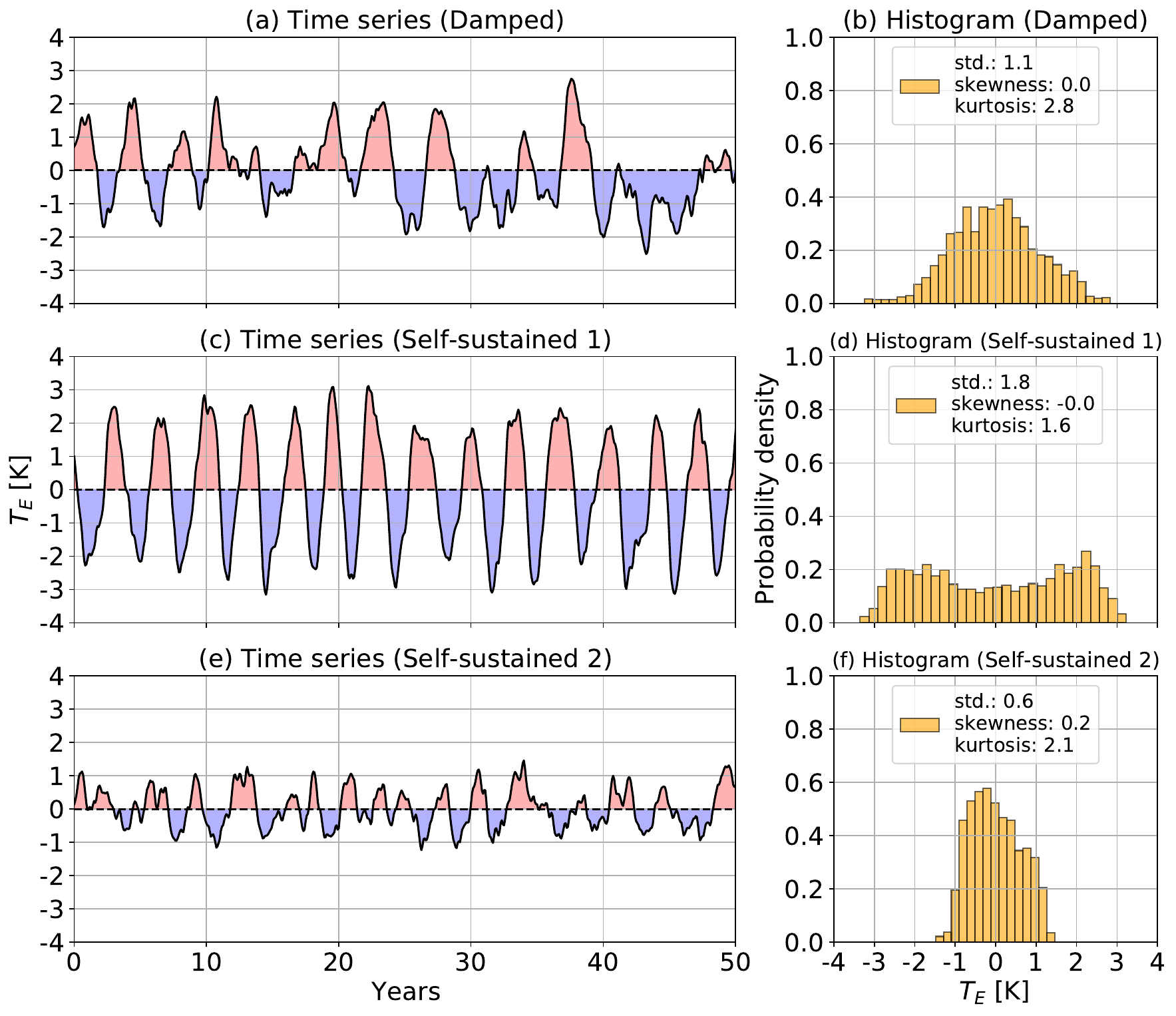}
  \caption{Niño index time series and the corresponding histograms for RO simulations comparing damped and self-sustained ENSO regimes. Panels (a) and (b) show results under damped regimes (Damped); panels (c) and (d) correspond to the idealized self-sustained regimes (Self-sustained 1); and panels (e) and (f) present results under a realistic self-sustained regime following \cite{Weeks25} (Self-sustained 2). See Table~\ref{tab:unstable1} for details on the models and parameter values used. The observed time series and corresponding histogram are shown in Figure~\ref{fig:observations}(a) and \ref{fig:observations}(e), respectively. 
  }
  \label{fig:unstable1}
\end{figure}

Figure~\ref{fig:unstable1} presents time series and corresponding histograms derived from the final 50 years of 150-year RO simulations for each simulation case. Even without detailed analysis, it is evident that the time series in the Damped case more closely resembles observations than the two other self-sustained cases. The data histogram for the Damped case exhibits an almost normal distribution, characterized by a kurtosis value of roughly 3.0\footnote{Here, we use kurtosis in the statistical sense, not to be confused with excess kurtosis, which is defined as kurtosis minus 3. Kurtosis is defined as the normalized fourth central moment: $\text{kurtosis}(T_E) = \frac{\mathbb{E}[(T_E - \mu)^4]}{\sigma^4}$, where $\mu$ and $\sigma$ are the mean and standard deviation of the variable $T_E$, respectively.}, which is typical of a Gaussian distribution. In contrast, the Self-sustained 1 case exhibits more regular cyclic behavior and a clear bimodal structure, resembling sine-like oscillations with a kurtosis value near 1.5. The time series and data distribution in the Self-sustained 2 case exhibits unique characteristics. Notably, there is a saturation of peak amplitudes around $\pm$~1~K. The corresponding kurtosis value is 2.1—intermediate between that of a bimodal distribution (1.5) and a normal distribution (3.0).

These behaviors can be interpreted as follows. In the Damped case, the system is inherently stable ($BJ < 0$), drawing data points toward zero. Stochastic forcing then perturbs these points, spreading them symmetrically around the origin and resulting in a distribution resembling a Gaussian shape. In contrast, in the Self-sustained Case 1 ($BJ > 0$), the system dynamics push data points away from zero. These outward tendencies are counteracted by the stabilizing effect of the cubic nonlinear term, $-c_T T_E^3$, which confines the data near the outer edges. In these regions, additional stochastic perturbations further disperse the data, leading to an accumulation near the boundaries and producing a bimodal distribution. In the Self-sustained Case 2, also with $BJ > 0$, the system includes much stronger cubic nonlinearity than in Case 1 (cf. $c_T = 0.088$ vs. $0.020$), effectively limiting the range of data and concentrating values closer to the center. As a result, the histogram appears to exhibit a single peak. However, the relatively low kurtosis value of 2.1 suggests underlying bimodal characteristics despite the unimodal appearance.

\begin{figure}[htbp]
  \centering
  \includegraphics[width=1.0\textwidth]{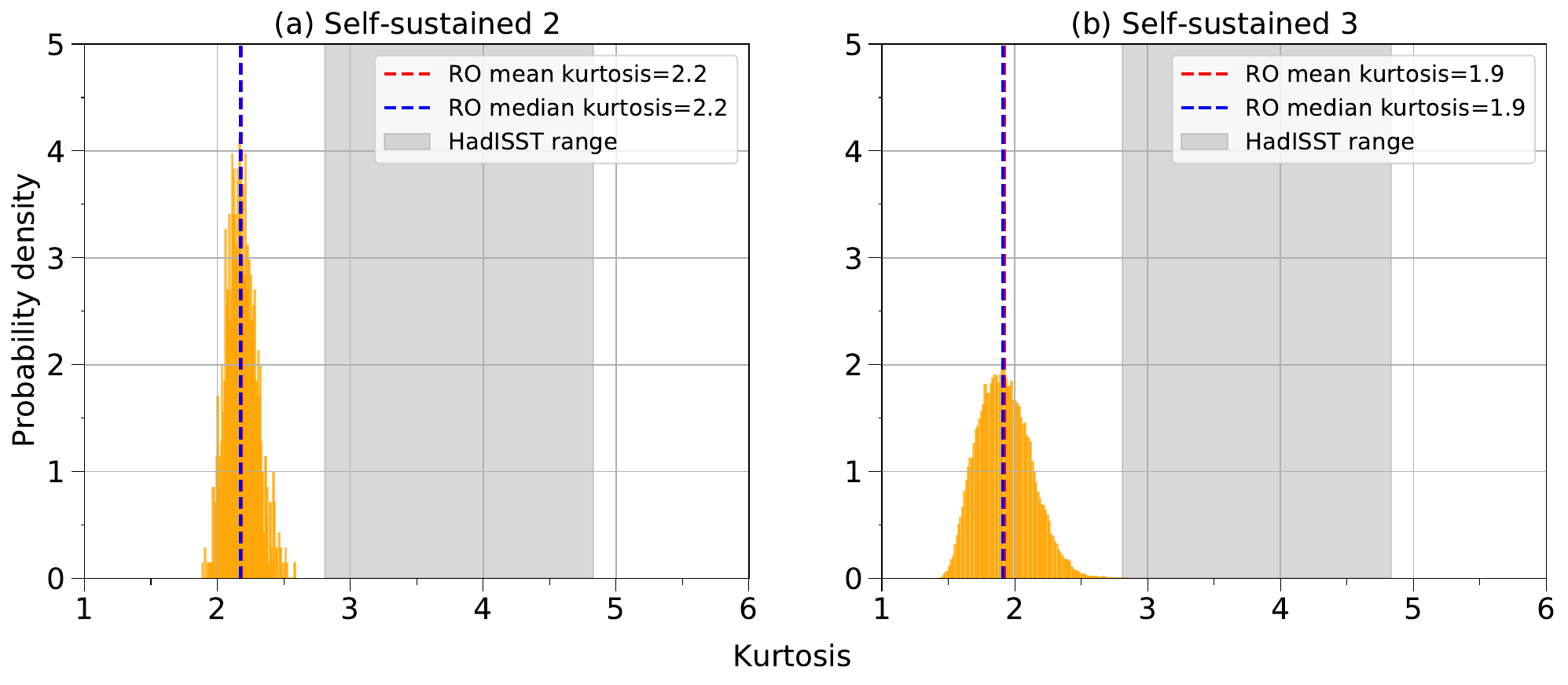}
  \caption{Distribution of kurtosis under self-sustained regimes from RO simulations using (a) the equation system and parameters from \citet{Weeks25}, modified to include white noise forcing (Self-sustained~2; Table~\ref{tab:unstable1}), and (b) parameter sweep experiments under the self-sustained regime (Self-sustained~3; Table~\ref{tab:unstable_parameter_sweeps}). Panel (a) is based on 1,000 time series, and panel (b) on 600 time series, each representing the final 50 years of a 150-year simulation. The observed kurtosis range, based on the Niño 3 index using a 50-year moving window advanced in 5-year increments over 1870–2020, is 2.8–4.8; the corresponding range for the Niño 3.4 index is 2.8–3.4 (not shown). See Figure~\ref{fig:best_statistics}(c) for comparison with the damped regime under the optimal case (Table~\ref{tab:parameter_sweep}).}
  \label{fig:unstable2}
\end{figure}

To further confirm the low-kurtosis behavior of the self-sustained system, we conducted additional experiments for Self-sustained Case 2, generating 1,000 time series using different realizations of stochastic forcing. Figure~\ref{fig:unstable2}(a) shows the distribution of kurtosis values calculated from the final 50 years of each 150-year RO simulation under this self-sustained configuration. The resulting mean and median kurtosis values are both 2.2—substantially lower than the observed range of 2.8–4.8 for the Niño 3 index (shown) and 2.8–3.4 for the Niño 3.4 index (not shown), based on 50-year moving windows (advanced in 5-year increments) over the period 1870–2020. These results reinforce our earlier conclusion that self-sustained systems tend to produce unrealistically low kurtosis values.

To assess whether this behavior generalizes beyond the specific configuration of \citet{Weeks25}, we performed another additional experiment involving a parameter sweep under the self-sustained regime (Self-sustained~3; see Table~\ref{tab:unstable_parameter_sweeps}). The kurtosis distribution, based on 600 time series—each representing the final 50 years of a 150-year simulation for each parameter combination—is shown in Figure~\ref{fig:unstable2}(b), yielding even lower mean and median kurtosis values of 1.9. This further supports the conclusion that self-sustained RO configurations would fail to reproduce the higher-order statistical properties observed in ENSO records. In stark contrast to the self-sustained cases, the damped regime—exemplified by the optimal case detailed in Table~\ref{tab:parameter_sweep} and discussed in Section~\ref{sec:parameter_sweep}—exhibits a higher kurtosis distribution, with a mean of 5.0 and a median of 4.7 (Figure~\ref{fig:best_statistics}(c)), which is closer to the observational ranges.

While skewness and kurtosis describe different aspects of a distribution—namely, asymmetry and tail heaviness—they may exhibit empirical correlation due to underlying dynamics or shared sources of variability. In fact, the highest skewness (0.9) and kurtosis (4.8) values in the observational record appear together during the 1970–2020 period, whereas the lowest skewness (0.2) and kurtosis (2.8) appear together during the 1920–1970 period (not shown). This correlation arises because extreme El Niño events, which are rare, contribute simultaneously to increased skewness and kurtosis: strong El Niño events shift the distribution toward more positive values (increasing skewness) and enhance the heaviness of the distribution’s tail (increasing kurtosis). Thus far, our investigations have focused on systems exhibiting low skewness, with mean values of 0.2 and 0.0 for Self-sustained Cases 2 and 3, respectively (not shown), within the self-sustained regime. This raises a natural question: can our conclusions be extended to higher-skewness systems that better reflect the observations of the past 50 years (e.g., skewness $\approx$ 0.9 in Figure~\ref{fig:observations}(f))?

To address this question, we have conducted additional experiments by modifying the Self-sustained 3 configuration to include $B = 1.0$ and $b_T = 0.05$ (results not shown). The value of $B$ was selected based on the highest value used in previous studies \citep{Levine10, Chen20}, while this value for $b_T$ is approximately 2 to 3 times larger than those used in prior work \citep{An20fokker, Chen20, Kim20, Kim21}. These larger values are intended to induce the highest possible skewness within the self-sustained regime. However, our simulations indicate that high skewness values (e.g., 0.9) could not be reproduced. This limitation arises from the presence of the $c_T$ term, whose cubic damping effect becomes increasingly dominant at large amplitudes. This damping suppresses the occurrence of extreme events, which are crucial for producing high skewness and kurtosis—a behavior clearly illustrated in Figure~\ref{fig:unstable1}(e) and (f), where the saturation of ENSO amplitudes is evident. In other words, self-sustained systems consistently yield distributions with low skewness and low kurtosis, as seen in the Self-sustained 2 and 3 cases.

These findings suggest that kurtosis can serve as a sensitive metric for distinguishing between dynamical regimes, indicating that the ENSO system observed in nature is unlikely to operate in a self-sustained regime and is instead more consistent with a damped system.

\begin{table}[htbp]
\caption{Equation system and model parameters used in our parameter sweep experiments under self-sustained regimes (Self-sustained 3). The numbers inside the brackets (\{...\}) indicate the parameter values explored. The parameter ranges are chosen to be inclusive and extensive, encompassing those explored in previous studies (see Table~1). Note that $R$ is confined to positive values in our study as a consequence of using $h_w$ as the state variable (see Section~\ref{sec:heat_content}). Listed parameter values are based on normalized units. For each parameter combination, 600 different realizations of 150-year-long time series are generated, with the last 50 years used for statistical analysis.}
\begin{center}
\begin{tabular}{lc}

\hline \hline
\multicolumn{2}{c}{Equations} \\ 
\hline
\\
\multicolumn{2}{l}{\makecell[l]{
$\frac{dT_E}{dt} = RT_E + F_1 h -c_T{T_E}^3+\sigma_T w_T$ \\ \\
$\frac{dh}{dt} = -\varepsilon h - F_2 T_E + \sigma_h w_h$
}} 
\\ \\ \hline \hline

& Self-sustained 3 (Parameter space where $\frac{R - \varepsilon}{2} > 0$)    \\ \hline

$R$
& $\{0.00, 0.02, ..., 0.28, 0.30\}$    

\\ \hline

$\varepsilon$ 
& $\{0.00, 0.02, ..., 0.28, 0.30\}$   

\\ \hline

$F_1$
& $\{0.15, 0.20, 0.25\}$   

\\ \hline 

$F_2$
& $\{0.15, 0.20, 0.25\}$   

\\ \hline  

$c_T$
& $\{0.00, 0.02, ..., 0.18, 0.20\}$   

\\ \hline  

$\sigma_T$ 
& $\{0.15, 0.20, 0.25\}$  

\\ \hline  

$\sigma_h$ 
& $\{0.15, 0.20, 0.25\}$ 

\\ \hline \hline 

\end{tabular}
\end{center}
\label{tab:unstable_parameter_sweeps}
\end{table}


\section{White or Red Noise Forcing?}
\label{sec:white_or_red}

In this section, we explore RO simulations under both white and red noise forcing assumptions, with the simulation setups summarized in Table~\ref{tab:white_red_noise}. We consider three simulation cases: White Noise, Red Noise 1 (with a decorrelation time of 45 days, following \citet{Jin07, Levine10, Levine15, Levine17, Chen20}), and Red Noise 2 (with a decorrelation time of 5 days, following \citet{Izumo24}). All three cases share identical linear parameter values and noise amplitudes, which are set to match those used in the Damped case from Table~\ref{tab:unstable1}. 

\begin{table}[htbp]
\caption{Equation system and model parameters to compare white and red noise forcing assumptions in RO simulations. Listed parameter values are based on normalized units.}
\begin{center}
\begin{tabular}{lccc}

\hline \hline

\multicolumn{3}{c}{Equations} \\ 
\hline
\\
\multicolumn{3}{l}{\makecell[c]{
$\frac{dT_E}{dt} = RT_E + F_1 h +\sigma_T \xi_T$ \\ \\
$\frac{dh}{dt} = -\varepsilon h - F_2 T_E + \sigma_h \xi_h$ \\ \\ $R=0.10$, $\varepsilon=0.20$, $F_1=F_2=0.20$, $\sigma_T=\sigma_h=0.20$
}} 
\\ \\ \hline \hline

& $\xi_{i=T,h}$ & $m_{i=T,h}$
\\ \hline 

White Noise
& $\xi_i=w_i$ 
& - 

\\ \hline \\

Red Noise 1 
& $\frac{d\xi_i}{dt}=-m_i\xi_i+\sqrt{2m_i}w_i$
& 0.67 (45 days)

\\ \\ \hline \\

Red Noise 2 
& $\frac{d\xi_i}{dt}=-m_i\xi_i+\sqrt{2m_i}w_i$
& 6 (5 days)

\\ \\ \hline \hline

\end{tabular}
\end{center}
\label{tab:white_red_noise}
\end{table}

\begin{figure}[htbp]
  \centering
  \includegraphics[width=1.0\textwidth]{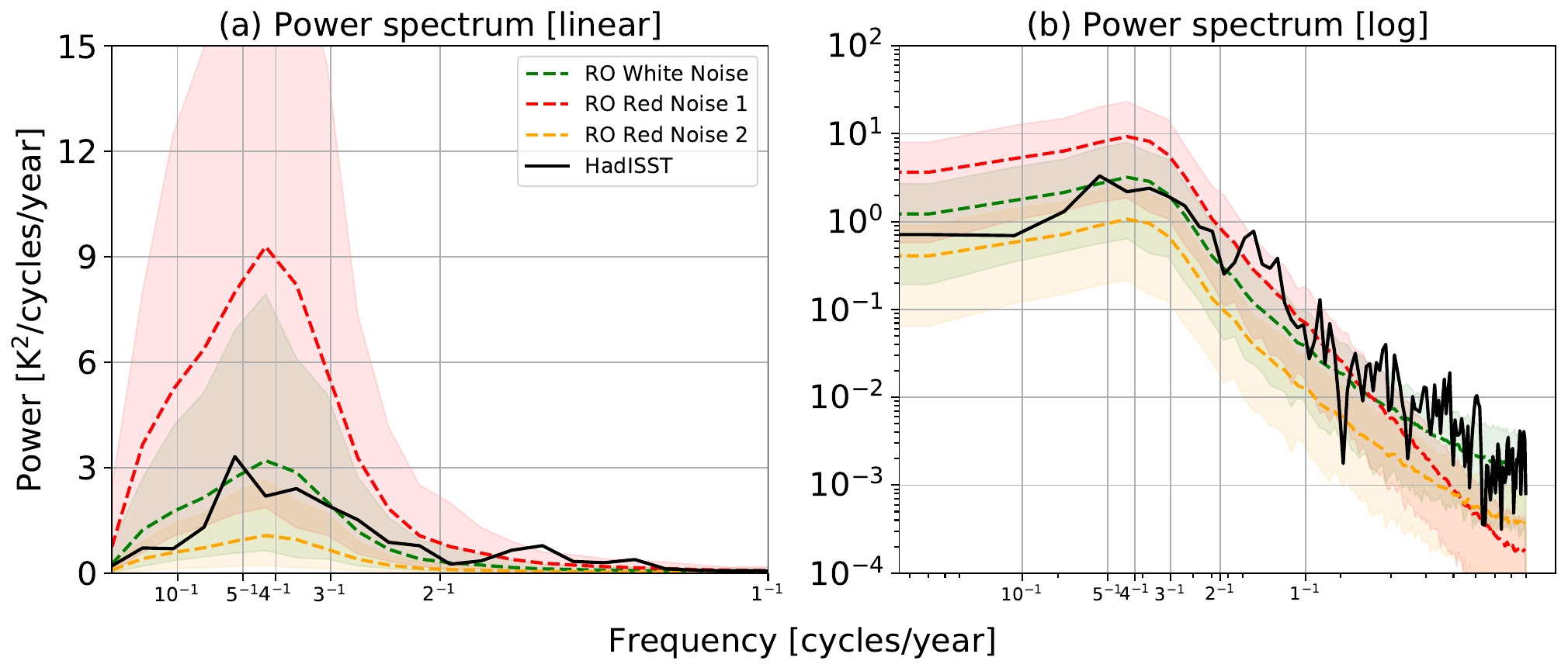}
  \caption{Power spectral density for RO simulations comparing white noise and two red noise forcing assumptions. For reference, the power spectrum computed from the HadISST Niño 3 index for the period 1970–2020 is shown as black curves. Simulations are run for 30,000 years and divided into 600 non-overlapping 50-year segments to compute the mean composite power spectrum (dashed line) and the 90\% confidence interval (shading). Panels (a) and (b) display the results in linear and logarithmic scales, respectively. See Table~\ref{tab:white_red_noise} for details on the adopted models and parameter values. 
  }
  \label{fig:power_spectrum}
\end{figure}

Figure~\ref{fig:power_spectrum} presents the power spectral density of various RO simulation cases, shown in both linear and logarithmic scales. The mean composite curves are computed from 600 non-overlapping 50-year segments derived from a 30,000-year simulation. While all simulations share identical model parameters except for the decorrelation time of the stochastic forcing, they exhibit distinct interannual (i.e., ENSO-related) peak strengths. Specifically, the Red noise 1 case shows the strongest spectral power, followed by the White noise case, and then the Red noise 2 case with the weakest. Furthermore, both the White noise and Red noise 2 cases display an intraseasonal-to-interannual spectral slope approximately proportional to $\omega^{-2}$, whereas the Red noise 1 case exhibits a steeper slope, proportional to $\omega^{-n}$ with $n > 2$.

To understand these behaviors, we derived the analytical expressions for the power spectral density in the RO framework (See Appendix~C). Under white noise forcing assumptions:

\begin{linenomath*} 
\begin{equation} 
|\tilde{T}_E (\omega)|^2 = \frac{N_0}{2} \cdot \frac{ F_1^2 \sigma_h^2 + (\omega^2 + \varepsilon^2) \sigma_T^2 } { (-\omega^2 - R\varepsilon + F_1 F_2)^2 + \omega^2 (\varepsilon - R)^2 } 
\end{equation} 
\end{linenomath*}

\noindent and under red noise forcing assumptions:

\begin{linenomath*}
\begin{equation}
|\tilde{T}_E (\omega)|^2 = \frac{N_0}{2} \cdot
\frac{ F_1^2 \sigma_h^2 \frac{2}{m_h\left(\frac{\omega^2}{m_h^2} + 1\right)} 
+ (\omega^2 + \varepsilon^2) \sigma_T^2 \frac{2}{m_T\left(\frac{\omega^2}{m_T^2} + 1\right)} }
{ (-\omega^2 - R\varepsilon + F_1 F_2)^2 + \omega^2 (\varepsilon - R)^2 }
\end{equation}
\end{linenomath*}

\noindent where $\frac{N_0}{2}$ is a normalization factor that ensures the total power is consistent across frequencies, accounting for the constant power distribution characteristic of white noise. The division by 2 reflects the distinction between the one-sided power spectrum (covering frequencies from 0 to $\infty$) and the two-sided spectrum (spanning $-\infty$ to $\infty$).

Under red noise forcing assumptions, the power spectra include additional multiplicative terms of the form $\frac{2}{m_{i=T,h} \left( \frac{\omega^2}{m_{i=T,h}^2} + 1 \right)}$ in the numerator. These terms modulate the frequency response and contribute to the varying $\omega^{-n}$ spectral slopes observed across different cases—for example, a steeper slope in the Red noise 1 case. For regions where $\omega \ll m_{i=T,h}$, this factor behaves approximately as $\approx\frac{2}{m_{i=T,h}}$. Thus, for the Red Noise 1 case, where $m_{i=T,h} = 0.67$, the scaling factor becomes 3, which implies roughly three times stronger energy concentration at low frequencies, including the ENSO spectral peak, compared to the White Noise case. For the Red Noise 2 case, where $m_{i=T,h} = 6$, the corresponding scaling factor is 0.33. This leads to approximately three times weaker energy concentration across all frequencies. The reduction in spectral power across all frequencies arises consistently when $m_{i=T,h} > 2$, as this results in the scaling factor $\frac{2}{m_{i=T,h}\left(\frac{\omega^2}{m_{i=T,h}^2} + 1\right)} < 1$, which uniformly attenuates power across the frequency domain.

For ENSO, variability in the interannual range dominates the power spectrum, while intraseasonal (high-frequency) and interdecadal (low-frequency) components play only a secondary role. In this sense, obtaining realistic ENSO amplitudes largely depends on matching the observed power in the interannual band. This can be achieved by adjusting the noise amplitudes—e.g., using three times smaller $\sigma_T$ and $\sigma_h$ in the Red noise 1 case, or three times larger values in the Red noise 2 case—or by applying a cubic nonlinearity term $c_T$, which can suppress the overall spectral power, as demonstrated in \cite{Chen20}. As we will detail later, Figures~\ref{fig:chen_simulation}(b) and (c) show the realistic interannual power peak that emerges from red noise forcing with a 45-day decorrelation time combined with a nonzero $c_T$. In the absence of $c_T$, the power spectrum resembles that of the Red noise 1 case, exhibiting excess energy in the low-frequency range (not shown). In other words, a realistic ENSO peak can still be achieved by adjusting model parameters. However, such adjustments require additional tuning without offering clear advantages. Therefore, the white noise forcing assumption is adopted hereafter as the simplest and most practical choice, which also allows replicating the observed power spectrum at low and high frequencies.

\section{The Role of Deterministic Nonlinearities and Multiplicative Noise}
\label{sec:nonlinear_parameters}

In this section, we aim to qualitatively examine the individual effects of each nonlinear term, before systematically evaluating which combination produces the most realistic representation of ENSO in the Section~\ref{sec:parameter_sweep}. Table~\ref{tab:nonlinear_comparisons} outlines the simulation setups used in this study. We consider five distinct cases: Linear, $B$ only, $b_T$ only, $d_T$ only, and $b_h$ only. All cases share identical linear parameter values and noise amplitudes, which are set to match those used in the Damped case from the Table~\ref{tab:unstable1}. Note that, unlike the other nonlinear parameters, $c_T$ does not contribute to ENSO amplitude, duration, or transition asymmetries. Rather, it is introduced as a means to stabilize the time series and will not be discussed in this section.

\begin{table}[htbp]
\caption{Equation system and model parameters to explore nonlinear parameters. Listed parameter values are based on normalized units.}
\begin{center}
\begin{tabular}{lc}

\hline \hline
\multicolumn{2}{c}{Equations} \\ 
\hline
\\
\multicolumn{2}{l}{\makecell[c]{
$\frac{dT_E}{dt} = RT_E + F_1 h + b_TT_E^2 + d_TT_Eh +\sigma_T w_T(1+BH(T)T)$ \\ \\
$\frac{dh}{dt} = -\varepsilon h - F_2 T_E -b_hT_E^2 + \sigma_h w_h$ \\ \\ $R=0.10$, $\varepsilon=0.20$, $F_1=F_2=0.20$, $\sigma_T=\sigma_h=0.20$
}} 
\\ \\ \hline \hline


& Nonlinear term \\ \hline

Linear
& All zero 

\\ \hline 

$B$ only
& $B=0.5$

\\ \hline 

$b_T$ only
& $b_T=0.015$

\\ \hline 

$d_T$ only
& $d_T=-0.015$

\\ \hline  

$b_h$ only
& $b_h=0.015$

\\ \hline \hline  

\end{tabular}
\end{center}
\label{tab:nonlinear_comparisons}
\end{table}

The parameter $B$ is set to 0.5, consistent with observational estimates \citep{Kug08, Levine17}. The deterministic nonlinear parameters $b_T$, $d_T$, and $b_h$ are set to 0.015, –0.015, and 0.015, respectively—adjusted from their initial regression-based estimates of 0.021, –0.017, and 0.013 to standardize their magnitudes. The chosen value of $b_T = 0.015$ is consistent with previous studies \citep{An20fokker, Chen20, Kim20, Kim21}, while the selected magnitude of $d_T = -0.015$ is slightly larger than earlier estimates \citep{An20fokker, Kim20}. Notably, our study adopts a negative sign for $d_T$, in contrast to the positive sign used in \citet{An20fokker} and \citet{Kim20}. This difference stems from the choice of heat content variable: we adopt $h = h_w$, whereas their studies use $h = h_{eq}$, resulting in a different phase relationship between $T_E$ and $h$ (see Figure~\ref{fig:phase_diagrams}). To our knowledge, a specific value for $b_h$ has not been previously reported.

\begin{figure}[htbp]
  \centering
  \includegraphics[width=0.43\textwidth]{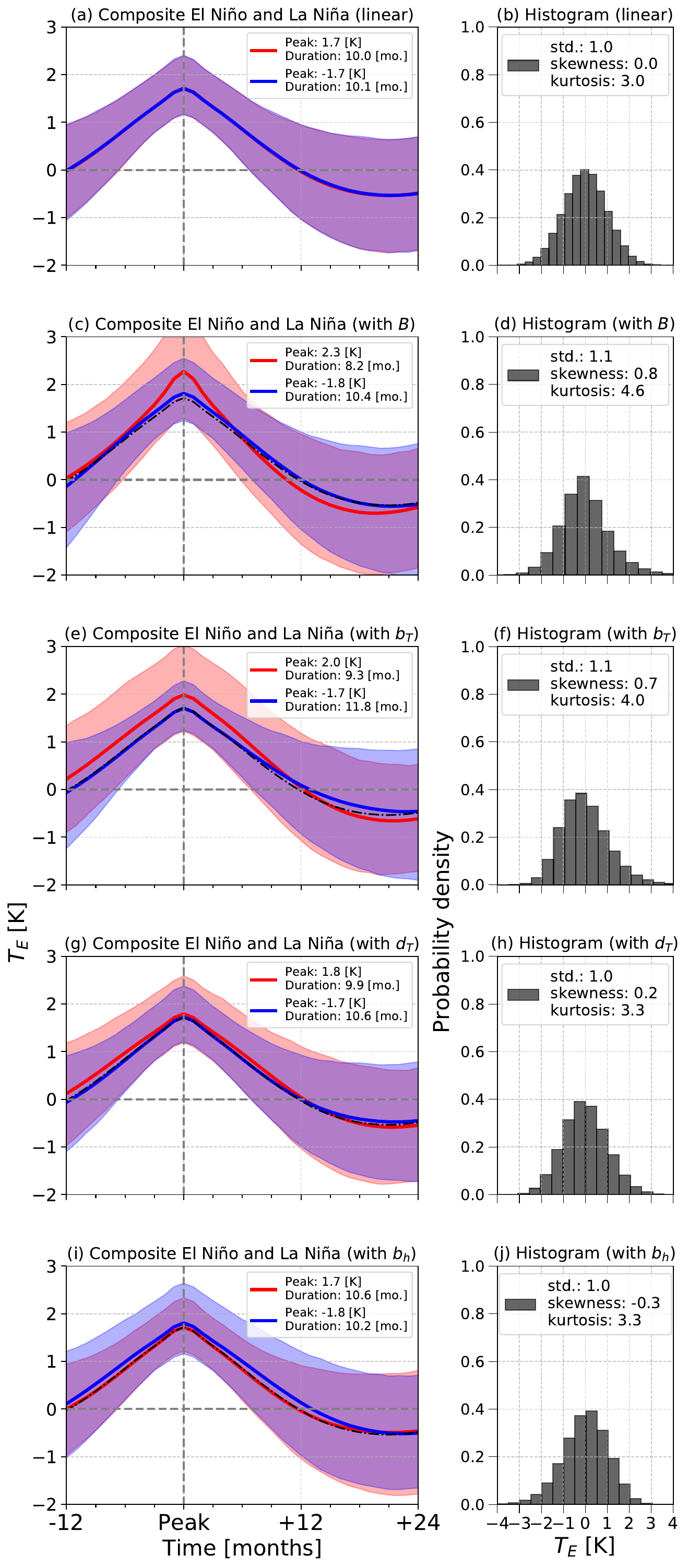}
  \caption{Composites of El Niño and La Niña temperature evolution, and data histogram for nonlinear test simulations described in Table~\ref{tab:nonlinear_comparisons}. Each panel represents a specific case: (a)-(b) Linear, (c)-(d) $B$ only, (e)-(f) $b_T$ only, (g)-(h) $d_T$ only, and (i)-(j) $b_h$ only. Results are based on 30,000 years of RO simulation data. Red and blue curves correspond to El Niño and La Niña events, respectively, with thick solid lines representing the mean composites. Shading indicates the 90\% confidence intervals. To highlight El Niño–La Niña asymmetry, the La Niña composite curve is shown with its sign reversed. The mean curve for the Linear case is shown as a dashed line in panels (c), (e), (g), and (i) for reference. Other notations used in this figure are consistent with those presented in Figure~\ref{fig:observations}.
  }
  \label{fig:NL_comparisons}
\end{figure}

Figure~\ref{fig:NL_comparisons} presents composite El Niño and La Niña evolutions, along with data histograms for each simulation case. To facilitate direct comparison of asymmetries, the La Niña curves have been sign-flipped. As it should, the Linear case exhibits no asymmetry: the El Niño composite evolution is a mirror image to that of La Niña where amplitude and duration of El Niño and La Niña events are symmetric (Figure~\ref{fig:NL_comparisons}(a)), and the skewness is near zero (Figure~\ref{fig:NL_comparisons}(c)). 

The $B$-only case reveals pronounced asymmetries in both amplitude and duration. El Niño events exhibit enhanced amplitudes—by approximately 0.5~K—and are shortened in duration by about two months (Figure~\ref{fig:NL_comparisons}(c)), compared to the Linear case. This results from the nature of multiplicative noise, which preferentially amplifies variability during warm phases. In contrast, La Niña durations remain largely similar to those in the Linear case, although a slight increase in La Niña strength by approximately 0.1~K is observed (Figure~\ref{fig:NL_comparisons}(c)).

The $b_T$-only case also induces amplitude and duration asymmetries, but through a different mechanism than the $B$-only case. In this case, the durations of both El Niño and La Niña events are affected—El Niño events become shorter, while La Niña events become longer (Figure~\ref{fig:NL_comparisons}(e) and (f)). When the temperature anomaly is defined such that the mean is zero, the skewness shifts the median below zero, resulting in more values being classified as negative anomalies. Consequently, La Niña events exhibit longer durations than El Niño events. The $d_T$ parameter exhibits qualitatively similar behavior to $b_T$ in terms of duration asymmetry, albeit with weaker effects (Figure~\ref{fig:NL_comparisons}(g) and (h)). This can be understood as follows: the term $d_T T_E h$ is approximately proportional to $-d_T T_E^2$, given the negative association between $T_E$ and $h$ for $h = h_w$ (Figure~\ref{fig:phase_diagrams}(a)). Thus, a negative $d_T$ effectively mimics the influence of the $b_T T_E^2$ term.

Finally, the $b_h$ parameter induces negative skewness in $T_E$ (Figure~\ref{fig:NL_comparisons}(j)). This contrasts with the other nonlinear terms examined, which produce positive skewness. This arises because the $-b_h T_E^2$ term in the $h$ equation systematically drives $h$ downward, resulting in negative skewness in $h$ (with $h_w$ exhibiting a skewness of –1.3). Through the $F_1 h$ term in the $T_E$ equation, this downward shift in $h$ subsequently suppresses $T_E$ as well. With the negative skewness in $T_E$, the $b_h$-only case produces stronger La Niña amplitudes than El Niño, which contrasts with observations (Figure~\ref{fig:NL_comparisons}(i)). However, the inclusion of $b_h$ is the only mechanism among the nonlinear effects considered that can account for the strong negative skewness observed in $h$ (\cite{Su09}; not shown).

Taken together, these results suggest that each nonlinear term contributes uniquely to ENSO asymmetries. In the next section, we explore how their combined effects give rise to the observed asymmetries.

\section{Exploring the Parameter Space to Simulate a Realistic ENSO}\label{sec:parameter_sweep}
\subsection{Method}

In this section, we aim to identify the regions of the RO parameter space that can reproduce the observed ENSO characteristics shown in Figure~\ref{fig:observations}. The nonlinear terms considered for generating ENSO asymmetries are $b_T$, $B$, and $b_h$, while $d_T$ is excluded. This decision is based on several considerations. First, as noted in the previous section, $d_T$ produces qualitatively similar behavior to $b_T$. Second, when $b_T$ and $d_T$ are estimated simultaneously through regression fitting, their estimated values vary significantly and are prone to sign flipping, as also reported by \citet{Kim20}. This is likely a result of multicollinearity between the two terms. Lastly, recent studies found that atmospheric nonlinearities play a more dominant role than oceanic nonlinearities in generating ENSO asymmetries \citep{Liu24, Srinivas24, Stuivenvolt-Allen25b}, and $d_T$ represents one such oceanic term. We also incorporate seasonality in $R$, as previous studies have shown that it explains ENSO seasonal synchronization \citep{Chen20, Kim21,Vialard25}. The cubic nonlinear term $c_T$ is retained in our study to ensure numerical stability in the simulations.

We first obtain baseline parameter values along with their error ranges (90\% confidence intervals) using regression fitting, with the exception of $B$. We then explore parameter combinations that extend well beyond these uncertainty ranges to ensure broad coverage of plausible values; the sampled ranges encompass those used in previous studies (Table~1). For $B$, we set an upper limit of $B = 1.0$ in our simulations, which is close to the $B = 0.9$ value adopted by \cite{Chen20}. It has been shown that as long as $B$ does not exceed 1.0, the standard deviation of the Niño index remains within the observed range \citep{Levine10}. Baseline values for $\sigma_T$ and $\sigma_h$ are obtained from the residuals of the regression fitting.

We run simulations for 30,000 years for each parameter combination and divide the resulting time series into 600 non-overlapping 50-year chunks. From these, we select chunks whose overall standard deviation, skewness, and kurtosis fall within the ranges 0.7–0.9, 0.2–0.9, and 2.8–4.8, respectively. These thresholds correspond to the 90\% confidence intervals derived from twenty different 50-year observational chunks spanning 1870–2020, using a 5-year moving average. Among the selected chunks, we further retain only those whose lagged autocorrelation structures and seasonal standard deviations also fall within the corresponding 90\% observational confidence intervals, computed using the same method. A parameter combination is identified as optimal if it yields the largest number of chunks that simultaneously satisfy all of these criteria. Table~\ref{tab:parameter_sweep} outlines our simulation setups and identified optimal parameter values.

\begin{table}[htbp]
\caption{Summary of the RO model setup, including the governing equations, baseline parameter values obtained from regression fitting, the parameter space explored in simulations, and the optimal parameter values identified for reproducing realistic ENSO behavior. The numbers inside the brackets (\{...\}) indicate the parameter values explored. The parameter ranges are chosen to be inclusive and extensive, encompassing those explored in previous studies (see Table~1) as well as the parameters obtained through the regression fitting method in this study (second column of the table). Experiments for each parameter combination are run for 30,000 years. From each simulation, 600 non-overlapping 50-year chunks are extracted, and only those chunks whose overall standard deviation, skewness, kurtosis, lagged autocorrelations, and seasonal standard deviations fall within the 90\% confidence intervals derived from twenty 50-year observational chunks (spanning 1870–2020, using a 5-year moving average) are retained. A parameter combination is deemed optimal if it produces the largest number of chunks that simultaneously satisfy all criteria. Parameter sweeps and optimization are initially conducted in normalized units, with optimal values subsequently converted to physical units using \(\text{std}(T_E) = 0.88\) and \(\text{std}(h_w) = 8.0\) (see Appendix~A).}
\begin{center}
\begin{tabular}{lcccc}

\hline \hline
\multicolumn{5}{c}{Equations} \\ 
\hline
\\
\multicolumn{5}{l}{\makecell[l]{
$\frac{dT_E}{dt} = (R_0-R_a\cos(\omega_at-\phi))T_E + F_1 h + b_TT_E^2 - c_TT_E^3 +\sigma_T w_T(1+BH(T)T)$ \\ \\
$\frac{dh}{dt} = -\varepsilon h - F_2 T_E -b_hT_E^2 + \sigma_h w_h$ \\  
}} 
\\ \\ \hline \hline

& \makecell[c]{Estimations \\ (Normalized)} &  \makecell[c]{Parameter Space \\ (Normalized)} & \makecell[c]{Optimal \\ (Normalized)} &
\makecell[c]{Optimal \\ (Raw)}

\\ \hline

$R_0$
& $0.028 \pm 0.018$ & $\{0.00, 0.01, ..., 0.19, 0.20\}$ & 0.03 & 0.03

\\ \hline 

$R_a$ 
& $0.115 \pm 0.022$ & $\{0.00, 0.04, ..., 0.16, 0.20\}$ & 0.16 & 0.16

\\ \hline

$\varepsilon$ 
& $0.196 \pm 0.012$ & $\{0.01, 0.02, ..., 0.35, 0.36\}$ & 0.13 & 0.13

\\ \hline

$F_1$
& $0.143 \pm 0.015$ & $\{0.10, 0.12, ..., 0.22, 0.24\}$ & 0.14 & 0.015

\\ \hline 

$F_2$
& $0.208 \pm 0.012$ & $\{0.10, 0.12, ..., 0.22, 0.24\}$ & 0.16 & 1.45

\\ \hline  

$b_T$ 
& $0.024 \pm 0.008$ & $\{0.00, 0.01, ..., 0.04, 0.05\}$ & 0.02 & 0.023

\\ \hline  

$c_T$ 
& $0.000 \pm 0.004$ & $\{0.000, 0.001, ..., 0.004, 0.005\}$ & 0.001 & 0.001

\\ \hline  

$B$ 
& - & $\{0.0, 0.2, ..., 0.8, 1.0\}$ & 0.4 & 0.45

\\ \hline 

$b_h$ 
& $0.012 \pm 0.005$ & $\{0.00, 0.01, ..., 0.04, 0.05\}$ & 0.03 & 0.310

\\ \hline 

$\sigma_T$ 
& $0.269$ & $\{0.16, 0.18, ..., 0.26, 0.28\}$ & 0.20 & 0.176

\\ \hline  

$\sigma_h$ 
& $0.231$ & $\{0.16, 0.18, ..., 0.26, 0.28\}$ & 0.20 & 1.60

\\ \hline \hline

\end{tabular}
\end{center}
\label{tab:parameter_sweep}
\end{table}


\subsection{Results}

\begin{figure}[htbp]
\centering
\includegraphics[width=0.86\textwidth]{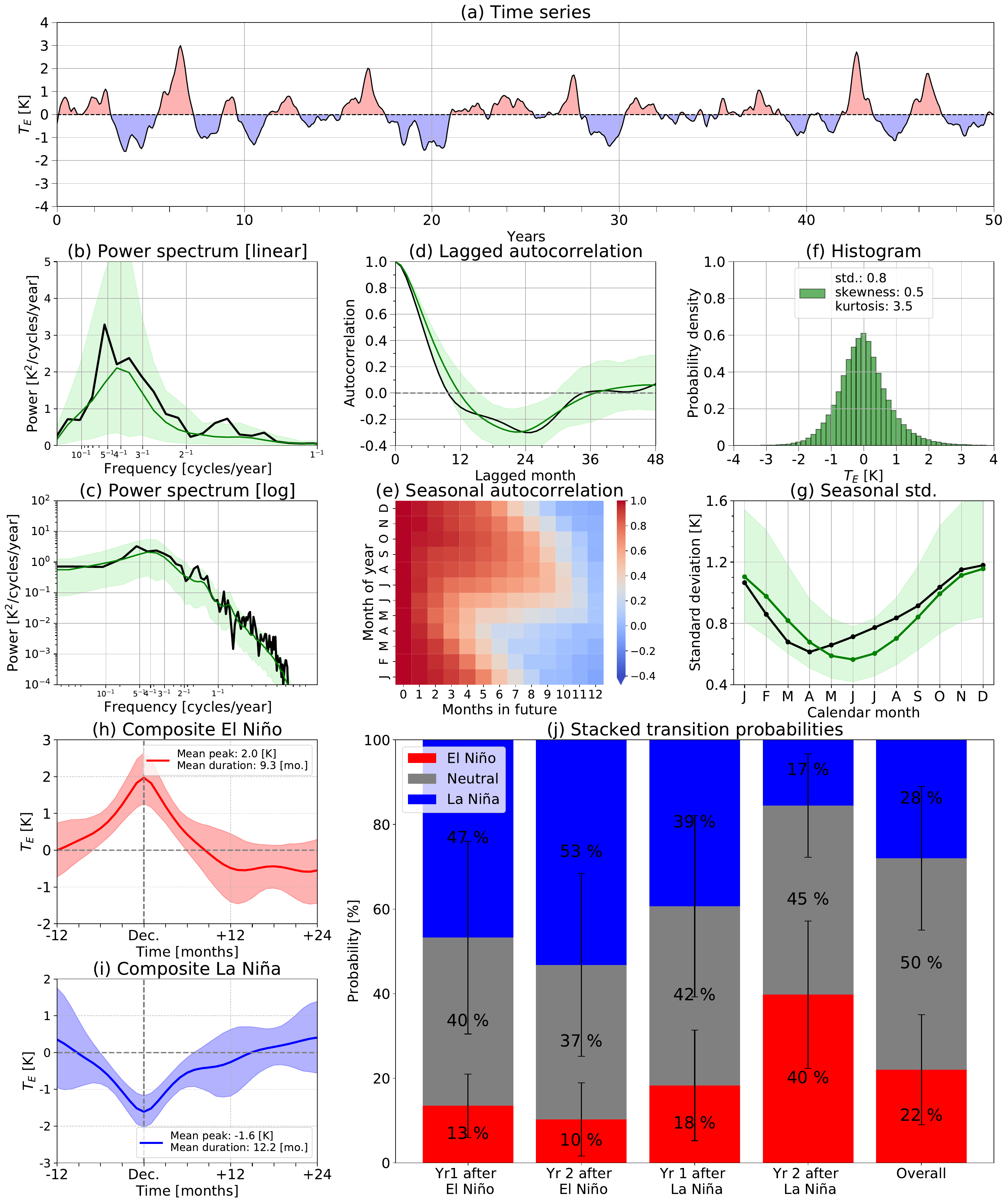}
\caption{As in Figure~\ref{fig:observations}, but based on RO simulations using the optimal parameters identified in this study (see Table~\ref{tab:parameter_sweep}). Mean composite curves and 90\% confidence intervals (shading) are shown for panels (b), (c), (d), (f), (g), (h), and (i), based on 600 non-overlapping 50-year segments. The confidence intervals shown as error bars in panel (j) represent the 90\% level. Panel (f) presents statistical moments computed from the full 30,000-year simulation, in contrast to Figure~\ref{fig:observations}(f), which is based on a single 50-year segment. For the distribution of statistical moments derived from the 600 segments, see Figure~\ref{fig:best_statistics}. We emphasize that here we aim to replicate ENSO characteristics as described by the Nino 3 index.}
\label{fig:best_simulation}
\end{figure}

A recent review highlights that state-of-the-art RO models can broadly reproduce several characteristics of ENSO, including its overall amplitude and amplitude asymmetry, the power spectrum, and seasonal phase locking \citep{Vialard25}—features that are also successfully reproduced in our study. Figure~\ref{fig:best_simulation}(h) and (i) show the composite evolution of moderate-to-strong El Niño and La Niña events, respectively, with El Niño exhibiting stronger amplitudes (2.0~K vs. -1.6~K). Figure~\ref{fig:best_simulation}(b) and (c) present the power spectral density on a linear and a logarithmic scale, revealing a broad interannual peak consistent with observations. Figure~\ref{fig:best_simulation}(e) shows that the annual cycle in $R$ is sufficient to reproduce the observed ENSO spring predictability barrier, consistent with the findings of \citet{Levine15}. Figure~\ref{fig:best_simulation}(g) demonstrates the seasonal locking of ENSO, with variability peaking in December. 

\begin{figure}[htbp]
  \centering
  \includegraphics[width=0.49\textwidth]{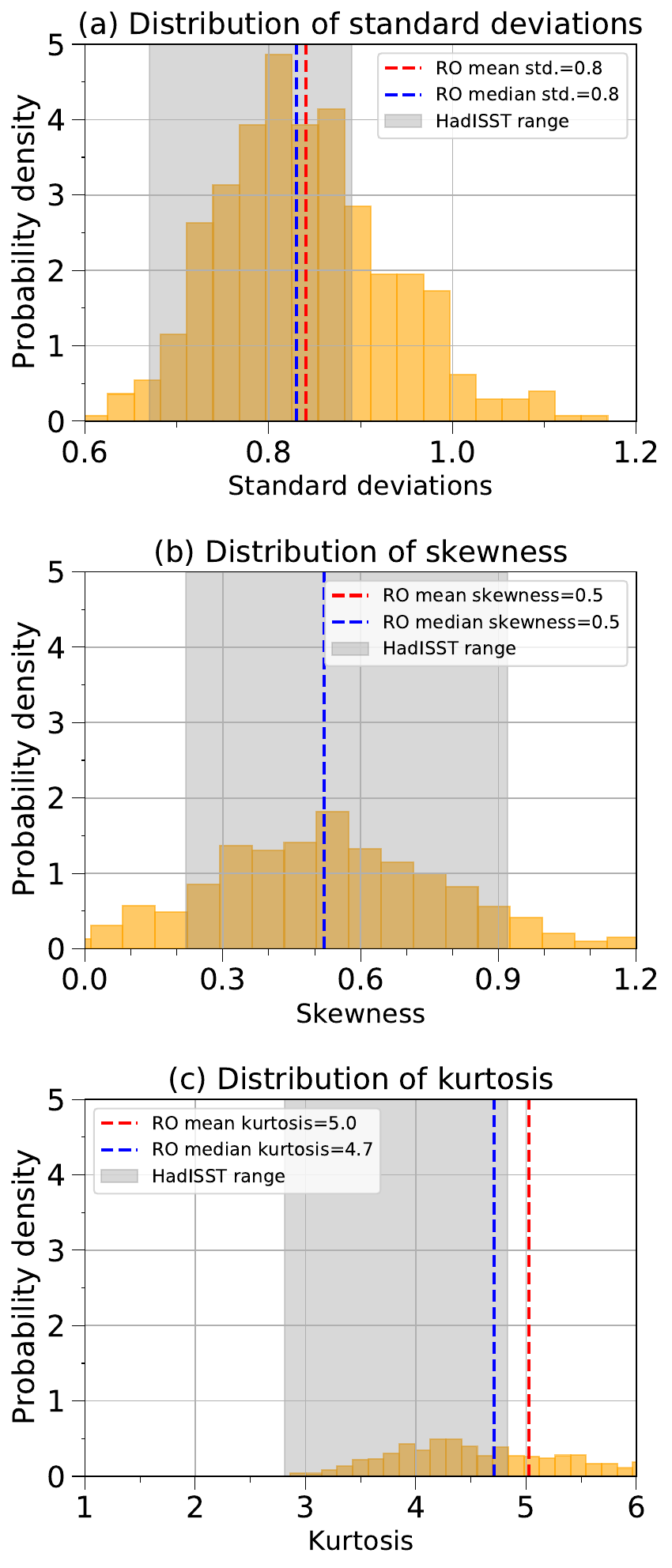}
  \caption{Distributions of standard deviation, skewness, and kurtosis computed from 600 non-overlapping 50-year segments extracted from a 30,000-year RO simulation using the optimal parameters and equation system summarized in Table~\ref{tab:parameter_sweep}. The 90\% confidence intervals of the observed ranges (shading) are computed from the Niño 3 index using a 50‑year moving window, advanced in 5‑year steps, for the period 1870–2020.}
  \label{fig:best_statistics}
\end{figure}

Figure~\ref{fig:best_simulation}(f) shows that the simulated standard deviation (0.8), skewness (0.5), and kurtosis (3.5) all lie within the corresponding observational ranges (0.7–0.9, 0.2–0.9, and 2.8–4.8, respectively). While the simulated skewness and kurtosis (0.5 and 3.5) appear lower than those in the observations (0.9 and 4.3)  (Figure~\ref{fig:observations}(f)), it is important to note that the former are calculated from a 30,000-year time series, whereas the latter are based on a specific 50-year segment that exhibits some of the highest skewness and kurtosis values among the 20 segments spanning 1870–2020. Figure~\ref{fig:best_statistics} shows the distributions of standard deviation, skewness, and kurtosis computed from 600 non-overlapping 50-year segments extracted from a 30,000-year optimal RO simulation. These distributions reveal substantial spread, with many segments exhibiting values comparable to or exceeding those in Figure~\ref{fig:observations}(f). Having established that our model reproduces the fundamental properties of ENSO, we now turn to less explored aspects: the lagged autocorrelation function, and asymmetries in event duration and phase transitions.

Figure~\ref{fig:best_simulation}(d) shows the lagged autocorrelation structure of the simulated Niño index, which exhibits a rapid decay within the first year followed by a more gradual decline, consistent with observations. The decay timescale of 20 months, corresponding to $BJ = -0.05$~month$^{-1}$ in our optimal simulation is significantly shorter than the dominant periodicity of around 50 months inferred from the autocorrelation structure and the power spectrum —indicating that the system is strongly damped. Moderate-to-strong El Niño events typically decay within a year, averaging 9.3 months in duration, whereas La Niña events persist longer at 12.2 months (Figures~\ref{fig:best_simulation}(h) and (i)), although this duration asymmetry is smaller than observations between 1970 and 2020 (7.0 months vs. 14.1 months). The La Niña probabilities at Year 1 and Year 2 following moderate-to-strong El Niño events are 47\% and 53\%, respectively—closely matching, though slightly lower than, the observed values of 50\% and 62\% (compare the first and second bars in Figure~\ref{fig:best_simulation}(j) with those in Figure~\ref{fig:observations}(j)). The La Niña probability one year after an initial moderate-to-strong La Niña event (i.e., the second-year La Niña probability) is somewhat lower in simulations than in observations (40\% vs. 56\%) (compare the third bar in Figure~\ref{fig:observations}(j) with that in Figure~\ref{fig:best_simulation}(j). The La Niña probability two years after an initial moderate-to-strong La Niña event (i.e., the third-year La Niña probability) is also lower in simulations than in observations (17\% vs 33\%) (compare the fourth bar in Figure~\ref{fig:best_simulation}(j) with that in Figure~\ref{fig:observations}(j)). Note again that the observational statistics are based on a specific 50-year segment with relatively high skewness (i.e., asymmetry), and should therefore be interpreted qualitatively when compared to our simulations, which use 600 non-overlapping 50-year segments with varying skewness drawn from a 30,000-year simulation. Notably, the 90\% confidence intervals shown in Figure~\ref{fig:observations}(j) are broad, and the observations fall well within the range of simulated transition probabilities.

\begin{figure}[htbp]
  \centering
  \includegraphics[width=0.93\textwidth]{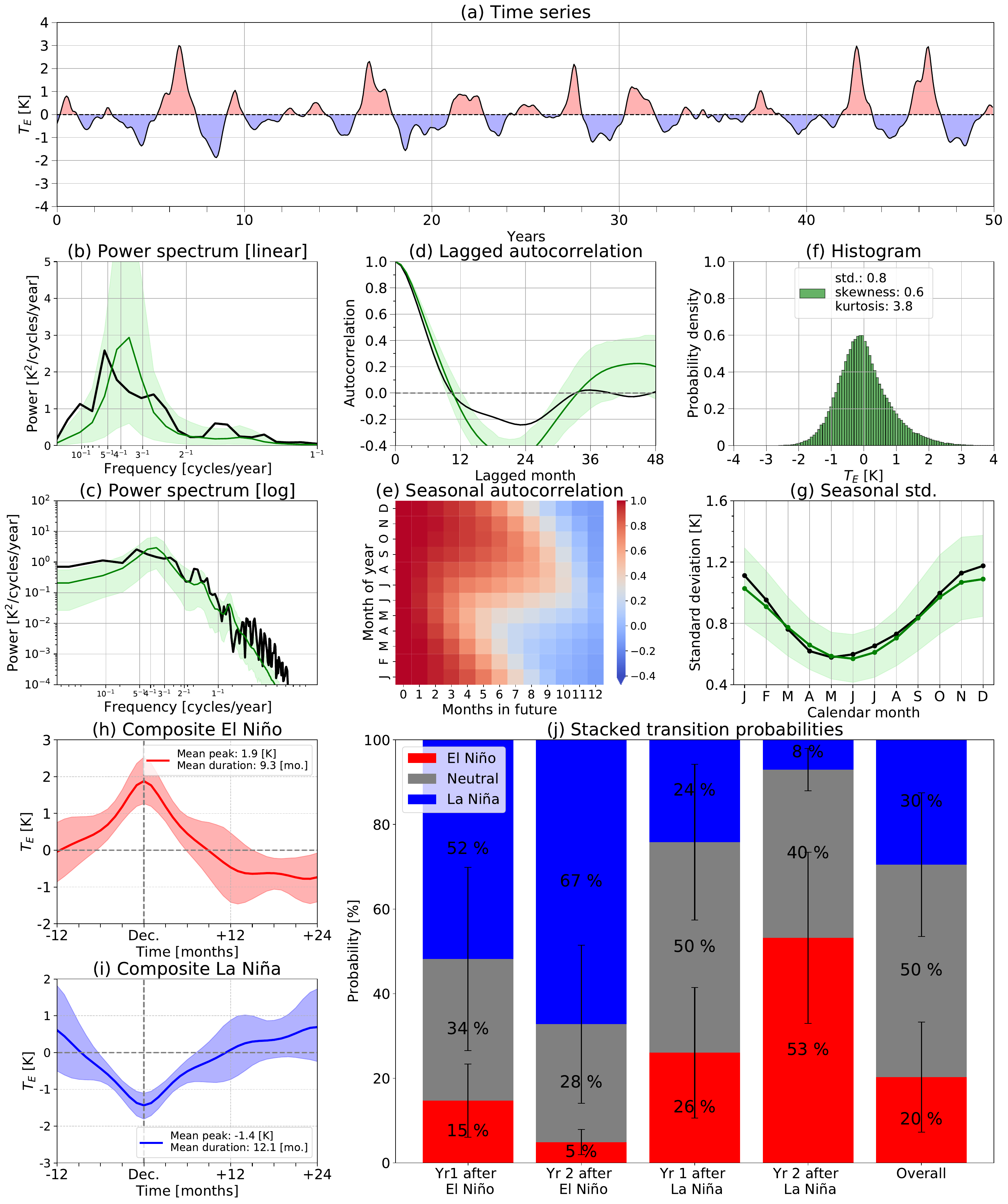}
  \caption{As in Figure~\ref{fig:best_simulation}, but based on RO simulations using the parameters from \cite{Chen20} (See Table~\ref{tab:chen}) to qualitatively compare their results with ours. For observational reference, the Niño 3.4 index (SSTA averaged over 170°W–120°W and 5°S–5°N) from HadISST data is used instead of the Niño 3 index, recognizing that \citet{Chen20} derived parameters based on the former.
  }
  \label{fig:chen_simulation}
\end{figure}

To demonstrate the improvements in reproducing key ENSO features—particularly asymmetries—we also conducted RO simulations using the parameter set from \cite{Chen20} (Table~\ref{tab:chen}) and evaluated the resulting ENSO characteristics based on these simulations (Figure~\ref{fig:chen_simulation}). \cite{Chen20} was chosen for comparison because they report parameter values necessary for simulations, and their parameter set also assumes $R > 0$ and $\varepsilon > 0$. The key difference between our optimal simulations and those of \citet{Chen20} is that we prescribe $b_h$, whereas \citet{Chen20} do not. Additionally, our study employs white noise forcing in both the $T_E$ and $h$ equations, while \citet{Chen20} apply red noise forcing to $T_E$ and no stochastic forcing to the $h$ equation. Several other parameter values also differ between the two studies, in part because our analysis is based on the Niño 3 index, whereas \citet{Chen20} use the Niño 3.4 index.

\begin{table}[htbp]
\caption{Equation system and model parameters from \citet{Chen20}, used to compare simulation results with those based on the optimal parameters. Based on the magnitudes of $F_1$ and $F_2$, we assume that the parameters are expressed in normalized units.}
\begin{center}
\begin{tabular}{lc}

\hline \hline
\multicolumn{2}{c}{Equations} \\ 
\hline
\\
\multicolumn{2}{l}{\makecell[l]{
$\frac{dT_E}{dt} = (R_0-R_a\cos(\omega_at-\phi))T_E + F_1 h + b_TT_E^2 - c_TT_E^3 +\sigma_T \xi_T(1+BH(T)T)$ \\ \\
$\frac{dh}{dt} = -\varepsilon h - F_2 T_E$ \\ \\
$\frac{d\xi_T}{dt} = -m_T \xi_T +w_T$ \\
}} 
\\ \\ \hline \hline

& Parameter Value    \\ \hline

$R_0$
& 0.043  

\\ \hline 

$R_a$ 
& 0.148  

\\ \hline

$\varepsilon$ 
& 0.087  

\\ \hline

$F_1$
& 0.146  

\\ \hline 

$F_2$
& 0.146  

\\ \hline  

$b_T$ 
& 0.018  

\\ \hline  

$c_T$ 
& 0.016  

\\ \hline  

$B$ 
& 0.9  

\\ \hline  

$\sigma_T$ 
& 0.111 

\\ \hline  

$m_T$ 
& 0.67 

\\ \hline \hline 

\end{tabular}
\end{center}
\label{tab:chen}
\end{table}

Comparisons between Figure~\ref{fig:best_simulation}(a) and Figure~\ref{fig:chen_simulation}(a) show that the simulated time series are qualitatively similar: El Niño events tend to be stronger and shorter-lived, while La Niña events are typically weaker but more persistent. The double- and triple-dip structure of La Niña events is also evident, highlighting their characteristic recurrence and extended duration. Nevertheless, differences become evident in other figures. For example, the red-noise formulation $\frac{d\xi_T}{dt} = -m_T \xi_T + w_T$ used in \citet{Chen20}, when combined with a strong $c_T$ term (16 times larger than in our optimal formulation), underrepresents power in both the low- and high-frequency domains, while the interannual peak remains consistent with the observations (Figure~\ref{fig:chen_simulation}(c)). The smaller $BJ \approx -0.022$~month$^{-1}$ value in their model—corresponding to a damping time scale of approximately 45 months—compared to $BJ = -0.05$~month$^{-1}$ (20 months) in our optimized simulations, is reflected in the lagged autocorrelation structure (Figure~\ref{fig:chen_simulation}(d)), which exhibits a less damped and more regular pattern than both the observations and our simulations.

The composite evolutions, as well as the amplitude and duration asymmetries, appear broadly similar between the two simulations (compare Figures~\ref{fig:best_simulation}(h) and (i) with Figures~\ref{fig:chen_simulation}(h) and (i)). However, notable differences emerge in the transition probabilities. La Niña probabilities remain relatively consistent across Year 1 and Year 2 following moderate-to-strong El Niño events, and Year 1 following moderate-to-strong La Niña events, both in observations (50\% and 62\% vs. 56\%) and in our optimal simulations (47\% and 53\% vs. 40\%). In contrast, the corresponding probabilities in \citet{Chen20} show large discrepancies (52\% and 67\% vs. 24\%). This is likely related to the insufficient skewness in $h$ under the parameter set from \citet{Chen20}, which yields a value of –0.4 due to the absence of the $b_h$ term. By comparison, the observed skewness of $h_w$ is –1.3, and our optimized model produces a comparable value of –1.1.

Taken together, these results suggest that the optimal equation system and parameter set identified in our study enhance the reproduction of key linear and nonlinear features of observed ENSO behavior. However, some discrepancies remain—particularly in the second-year and third-year La Niña transition probability and the magnitude of duration asymmetry. If not due to incomplete parameter specification, these differences are likely attributable to the limited size of the observational sample. The historical record contains only about 10 El Niño events over the past 50 years, in contrast to approximately 5,000 events generated over 30,000 years in our RO simulations. Consequently, composite curves and transition probabilities estimated from observational periods with high skewness are more susceptible to sampling variability and potential bias.

\section{Discussion and Conclusions}
\label{sec:discussion_and_conclusions}

In this study, we showed that the choice of the heat content variable $h$ is intrinsically linked to the signs of $R$ and $\varepsilon$ through their sum, $R + \varepsilon$, as well as to its phase relationship with the temperature anomaly. Relatedly, the sign of $R$ varies depending on the chosen heat content variable as shown in Table~1. This might suggest that subprocesses contributing to the Bjerknes feedback coefficient, $R$—including zonal advective feedback, Ekman upwelling feedback, and thermocline feedback—as well as dynamic damping by mean currents and thermodynamic damping, have differing impacts depending on the definition of $h$ \citep{Kim11a, Kim14, Jin20review}. In particular, damping processes appear to dominate when $h = h_{eq}$, resulting in a negative $R$, while positive feedback processes dominate when $h = h_w$, yielding a positive $R$. A more detailed and quantitative investigation is warranted to further explore these dependencies, not only for $R$ but also for $\varepsilon$.

We showed that self-sustained ENSO regimes produce unrealistically low kurtosis compared to observations, due to the inherent regularity of their oscillatory cycles. We propose that kurtosis serves as an effective diagnostic for distinguishing between damped and self-sustained dynamical regimes. Our parameter sweep simulations and kurtosis analysis further support the conclusion that ENSO is best characterized as a damped oscillator within the RO framework.

In the context of the noise structure, near-monthly decorrelation timescales have previously been inferred from the autocorrelation of residuals in regression-based model fitting \citep{Jin07, Levine10, Levine15, Levine17}. However, adding red noise to RO simulations increases low-frequency power and suppresses high-frequency power compared to white noise forcing. While the interannual ENSO power peak can be reproduced with either white or red noise forcing, explicitly adding red noise introduces unnecessary complexity without clear benefits. White noise forcing alone is sufficient to capture the spectral characteristics across both high and low frequencies. Relatedly, one should note that residual autocorrelation is not necessarily evidence of red noise; more often, it reflects model deficiencies that produce structured, non-white residuals.

In our exploration of nonlinear terms, we found that the multiplicative noise term $B$ and the deterministic nonlinear terms $b_T$ and $b_h$ act to enhance ENSO asymmetries. Physically, $B$ represents the positive feedback between WWBs and SSTAs, linking amplitude and duration asymmetries by enabling strong post–El Niño cooling that overshoots the neutral state, consistent with the mechanism proposed by \citet{Choi13} in the DO framework. The terms $b_T$ and $b_h$ arise naturally when wind stress is assumed to depend nonlinearly on temperature anomalies \citep{Jin20review}. In particular, the previously underemphasized $b_h$ term plays a critical role in generating sufficient negative skewness in $h$, which is key to reproducing the observed $T_E$ and $h$ phase relationship and sustaining multi-year La Niñas after moderate-to-strong El Niños. These findings highlight the central role of atmospheric nonlinearities in generating the observed ENSO asymmetries, while not discounting the contribution of oceanic nonlinearities.

Our optimal RO formulation incorporates seasonality in the Bjerknes feedback strength, includes both quadratic and cubic nonlinearities as well as multiplicative noise in the $T_E$ equation, and introduces a quadratic temperature term in the $h$ equation, all under the assumption of white noise forcing. The optimal parameter set yields a stable and damped system with $BJ = -0.05$~month$^{-1}$ and $Wyrtki = 0.126$~month$^{-1}$, while successfully reproducing key observational features, including the power spectral density, phase locking, spring predictability barrier, lagged autocorrelation structure, statistical moments (standard deviation, skewness, and kurtosis), and amplitude, duration, and phase transition asymmetries between El Niño and La Niña. The decay timescale of 20 months, corresponding to $BJ = -0.05$~month$^{-1}$, is shorter than ENSO’s dominant periodicity of approximately four years, and plays a key role in rendering the system more damped and less periodic \citep{Kessler02, Philander03}.

As shown in Figure~\ref{fig:best_statistics}, the statistical moments from the optimal formulation exhibit substantial spread across 50‑year segments, despite all time series being generated with the same parameters. This spread suggests the potential for large internal variability in the ENSO system. Consequently, the observed ENSO variations might arise solely from internal variability, even in the absence of prescribed parameter changes—and thus without changes in the mean background state or underlying physical processes \citep[e.g.,][]{Wittenberg09, Wittenberg14, Vega-Westhoff17, Maher18, Fedorov20review, Vialard25}. This question will be explored in detail in a future study.

This study adopted the Niño 3 index, which allows ENSO’s asymmetric characteristics to be more clearly distinguished than when using the Niño 3.4 index. Our optimal RO formulation is therefore biased toward Eastern Pacific (EP) rather than Central Pacific (CP) El Niño events. Several recent studies have incorporated not only the eastern Pacific temperature anomaly ($T_E$) but also the central Pacific one ($T_C$) into their RO frameworks \citep{Fang18, Geng20, Chen22} to account for CP events, thereby capturing the diversity of ENSO patterns \citep{Capotondi15, Timmermann18, Capotondi20review}. Including $T_C$ and/or using the Niño 3.4 index instead of the Niño 3 index may place different emphasis on the importance of specific nonlinear terms, potentially leading to different interpretations and conclusions \citep{Vialard25}. That said, the Niño 3.4 index exhibits a kurtosis value of approximately 3.0, which is characteristic of a normal distribution. Moreover, CP events are on average weaker than EP events. This suggests that one of our main conclusions—that ENSO operates in a damped regime—would likely remain unchanged.

In summary, this study presents a comprehensive investigation of the RO framework by systematically exploring linear and nonlinear parameters, and optimizing both equation structure and parameter values to reproduce observed ENSO characteristics. Our results provide new insights into the physical mechanisms driving ENSO asymmetries and demonstrate key improvements achievable through an optimal formulation. The resulting equation structure and parameters can serve as a baseline for realistic ENSO simulations within the RO framework and as a diagnostic tool for evaluating how key nonlinearities are represented in GCMs. These efforts ultimately contribute to improving the interpretability and predictability of ENSO and enhancing future climate projection capabilities.

\acknowledgments

The authors thank three anonymous reviewers and Eli Tziperman for their constructive comments. We also acknowledge the use of the ChatGPT-4o language model for assistance in improving the clarity and readability of the manuscript. Sooman Han thanks Bastien Pagli for providing data preprocessing tools and acknowledges Clara Deser and Soong-Ki Kim for valuable discussions that contributed to the development of this work. Alexey V. Fedorov has been supported by NOAA (NA20OAR4310377), DOE (DE-SC0023134), and by the ARCHANGE project (ANR-18-MPGA-0001, France). Jérôme Vialard has been supported by the ARISE project (ANR-18-CE01-0012).


%
%
\datastatement

The observational HadISST data used in this study \citep{Rayner03} are open-access and can be obtained from the following sources: Met Office Hadley Centre  (https://www.metoffice.gov.uk/hadobs/hadisst/) for global SST data, and NOAA Physical Sciences Laboratory (https://psl.noaa.gov/data/) for the Niño index data. The ORAS5 reanalysis data used in this study \citep{Zuo19} are also open-access and available through the Copernicus Climate Data Store (https://cds.climate.copernicus.eu). The Recharge Oscillator simulation code is archived on Zenodo as \citet{HanROcode25}.


%






%



\appendix[A]
\appendixtitle{Normalization of State Variables in the Recharge Oscillator}

We consider the following form of the nonlinear RO (c.f. \cite{Jin20review, Vialard25}):

\begin{linenomath*}
\begin{equation}
\frac{dT_E}{dt} = RT_E + F_1 h + b_T T_E^2 - c_T T_E^3 + d_T T_E h + \sigma_T w_T (1 + BH(T_E) T_E) 
\end{equation}

\begin{equation}
\frac{dh}{dt} = -\varepsilon h - F_2 T_E - b_h T_E^2 + \sigma_h w_h
\end{equation}
\end{linenomath*}

\noindent The variables $T_E$ and $h$ have units of K and m, respectively. The linear parameters $R$, $F_1$, $\varepsilon$, and $F_2$ have units of month$^{-1}$, K$\cdot$m$^{-1}\cdot$month$^{-1}$, month$^{-1}$, and m$\cdot$K$^{-1}\cdot$month$^{-1}$, respectively. The nonlinear parameters $b_T$, $c_T$, $d_T$, and $b_h$ are expressed in units of K$^{-1}\cdot$month$^{-1}$, K$^{-2}\cdot$month$^{-1}$, m$^{-1}\cdot$month$^{-1}$, and m$\cdot$K$^{-2}\cdot$month$^{-1}$, respectively. $w_T$ and $w_h$ are uncorrelated white noise time series with unit variance. Since the Wiener process has units of time$^{-0.5}$, the noise amplitudes $\sigma_T$ and $\sigma_h$ are interpreted to have units of K$\cdot$month$^{-0.5}$ and m$\cdot$month$^{-0.5}$, respectively. Multiplicative noise parameter $B$ has units of K$^{-1}$. The Heaviside function $H(T_E)$ is defined as:  

\begin{linenomath*}
\begin{equation}
H(T_E) =
\begin{cases}
1 & \text{if} \ T_E > 0 \\
0 & \text{if} \ T_E \leq 0
\end{cases}
\end{equation}
\end{linenomath*}

\noindent The Heaviside function is dimensionless and has no units.

To normalize the equations, we define the dimensionless variables $\hat{T}_E = \frac{T_E}{\text{std}(T_E)}$ and $\hat{h} = \frac{h}{\text{std}(h)}$. This transformation ensures that the new state variables, $\hat{T}_E$ and $\hat{h}$, have comparable scales, while also ensuring that the parameters maintain consistent units. The resulting normalized equations are as follows:

\begin{linenomath*}
\begin{equation}
\frac{d\hat{T}_E}{dt} = \hat{R} \hat{T}_E + \hat{F}_1 \hat{h} + \hat{b}_T \hat{T}_E^2 - \hat{c}_T \hat{T}_E^3 + \hat{d}_T \hat{T}_E \hat{h} 
+ \hat{\sigma}_T w_T (1 + \hat{B} H(\hat{T}_E)\hat{T}_E)
\end{equation}

\begin{equation}
\frac{d\hat{h}}{dt} = -\hat{\epsilon} \hat{h} - \hat{F}_2 \hat{T}_E - \hat{b}_h \hat{T}_E^2 + \hat{\sigma}_h w_h
\end{equation}
\end{linenomath*}

\noindent where the transformed parameters are defined as: 

\begin{linenomath*}
\begin{equation}
\hat{R} = R
\end{equation}

\begin{equation}
\hat{F}_1 = \frac{\text{std}(h)}{\text{std}(T_E)} F_1
\end{equation}

\begin{equation}
\hat{\varepsilon} = \varepsilon
\end{equation}

\begin{equation}
\hat{F}_2 = \frac{\text{std}(T_E)}{\text{std}(h)} F_2
\end{equation}

\begin{equation}
\hat{b}_T = \text{std}(T_E) b_T
\end{equation}

\begin{equation}
\hat{c}_T = \text{std}(T_E)^2 c_T
\end{equation}

\begin{equation}
\hat{d}_T = \text{std}(h) d_T
\end{equation}

\begin{equation}
\hat{b}_h = \frac{\text{std}(T_E)^2}{\text{std}(h)} b_h
\end{equation}

\begin{equation}
\hat{B} = B \cdot\text{std}(T_E)
\end{equation}

\begin{equation}
\hat{\sigma}_T = \frac{\sigma_T}{\text{std}(T_E)}
\end{equation}

\begin{equation}
\hat{\sigma}_h = \frac{\sigma_h}{\text{std}(h)}
\end{equation}
\end{linenomath*}

\noindent In this form, the variables $\hat{T}_E$, $\hat{h}$, and $\hat{B}$ are dimensionless, while the normalized linear parameters $\hat{R}$, $\hat{F}_1$, $\hat{\varepsilon}$, and $\hat{F}_2$, nonlinear parameters $\hat{b}_T$, $\hat{c}_T$, $\hat{d}_T$, and $\hat{b}_h$, have units of month$^{-1}$ and noise parameters $\hat{\sigma}_T$ and $\hat{\sigma}_h$ have units of month$^{-0.5}$. 

The expression for the indices do not change whether the original or normalized equations are used:

\begin{linenomath*}
\begin{equation}
BJ = \frac{R - \varepsilon}{2} = \frac{\hat{R} - \hat{\varepsilon}}{2}
\end{equation}

\begin{equation}
Wyrtki = \sqrt{ F_1 F_2 - \frac{(R + \varepsilon)^2}{4} } = \sqrt{ \hat{F}_1 \hat{F}_2 - \frac{(\hat{R} + \hat{\varepsilon})^2}{4} }
\end{equation}
\end{linenomath*}


\appendix[B]
\appendixtitle{Analytical Expressions for the Phase Space Diagram Slope}

We consider the simplest form of the linear RO \citep{Burgers05}:

\begin{linenomath*}
\begin{equation}
\frac{dT_E}{dt} = R T_E + F_1 h
\end{equation}

\begin{equation}
\frac{dh}{dt} = -\varepsilon h - F_2 T_E
\end{equation}
\end{linenomath*}

For initial conditions $(T_E(t=0),h(t=0))=(T_0, h_0)$, the analytical solutions are given as

\begin{linenomath*}
\begin{align}
T_E(t) &= T_0 \exp(BJ \cdot t) \cos(Wyrtki \cdot t) \nonumber \\
&\quad + \left[ \frac{\varepsilon + BJ}{Wyrtki} T_0 
+ \frac{(\varepsilon + BJ)^2 + Wyrtki^2}{F_2 \cdot Wyrtki} h_0 \right] 
\exp(BJ \cdot t) \sin(Wyrtki \cdot t)
\end{align}

\begin{equation}
h(t) = h_0 \exp(BJ \cdot t) \cos(Wyrtki \cdot t) 
- \frac{F_2 T_0 + (\varepsilon + BJ) h_0}{Wyrtki} 
\exp(BJ \cdot t) \sin(Wyrtki \cdot t)
\end{equation}
\end{linenomath*}

\noindent Without losing generality, the solutions can be simplified by choosing the initial conditions as $(T_0,h_0)=(T_0,0)$. Additionally, the exponential term $\exp(BJ \cdot t)$ can be omitted, as it acts as a scaling factor for amplitudes and does not affect the slope in the phase space diagram. The simplified solutions become:

\begin{linenomath*}
\begin{equation}
T_E(t) = T_0 \cos(Wyrtki \cdot t) 
+ \frac{\varepsilon + BJ}{Wyrtki} T_0 \sin(Wyrtki \cdot t)
\end{equation}

\begin{equation}
h(t) = -\frac{F_2 T_0}{Wyrtki} \sin(Wyrtki \cdot t)
\end{equation}
\end{linenomath*}

$T_E(t)$ and $h(t)$ can be expressed as cosine functions: 

\begin{linenomath*}
\begin{equation}
T_E(t) = R_T \cos(Wyrtki \cdot t + \varphi_T)
\end{equation}

\begin{equation}
h(t) = R_h \cos(Wyrtki \cdot t + \varphi_h)
\end{equation}
\end{linenomath*}

\noindent where the amplitudes and phase angles are given by:

\begin{linenomath*}
\begin{equation}
R_T^2 = T_0^2 + \left( \frac{\varepsilon + BJ}{Wyrtki} T_0 \right)^2
= \left( 1 + \frac{(\varepsilon + BJ)^2}{Wyrtki^2} \right) T_0^2
\end{equation}

\begin{equation}
R_h^2 = 0^2 + \left( -\frac{F_2 T_0}{Wyrtki} \right)^2
= \frac{F_2^2}{Wyrtki^2} T_0^2
\end{equation}

\begin{equation}
\varphi_T = \arctan \left( \frac{\frac{(\varepsilon + BJ)}{Wyrtki} \cdot T_0}{T_0} \right)
= \arctan \left( \frac{\varepsilon + BJ}{Wyrtki} \right)
\end{equation}

\begin{equation}
\varphi_h = \arctan \left( \frac{ - (F_2 T_0)/Wyrtki}{0} \right)
= \frac{\pi}{2}
\end{equation}
\end{linenomath*}

The slope $\theta$ in the phase space diagram is given by:

\begin{linenomath*}
\begin{equation}
\tan(2\theta) = \frac{2 R_T R_h \cos(\varphi_h - \varphi_T)}
{R_T^2 - R_h^2}
\end{equation}
\end{linenomath*}

\noindent By substituting $R_T$, $R_h$, $\varphi_T$, and $\varphi_h$, the expression for $\theta$ in terms of $R$, $\varepsilon$, $F_1$, and $F_2$ is:

\begin{linenomath*}
\begin{equation}
\tan(2\theta) = \frac{-(R + \varepsilon)}{F_1 - F_2}
\end{equation}
\end{linenomath*}

\noindent or equivalently,

\begin{linenomath*}
\begin{equation}
\theta = \frac{1}{2} \arctan\left(\frac{-(R + \varepsilon)}{F_1 - F_2} \right)
\end{equation}
\end{linenomath*}

It is important to note that both the numerator and denominator contain information about the angular quadrant, and preserving their respective signs is essential for correctly determining the phase. For example, $\frac{-(\hat{R} + \hat{\varepsilon})}{\hat{F_1} - \hat{F_2}}$ and $\frac{\hat{R} + \hat{\varepsilon}}{-(\hat{F_1} - \hat{F_2})}$ as an argument for arctangent should give a different signs for $\theta$. For the tangent function, the sign of $\theta$ depends only on the sign of the numerator, i.e., $-(\hat{R} + \hat{\varepsilon})$.\footnote{For example, given $\tan(\theta) = \frac{y}{x}$, consider four cases: $(x, y) = (1, 1),\, (-1, 1),\, (1, -1),\, (-1, -1)$, which correspond to $\theta = 45^\circ,\, 135^\circ,\, -45^\circ,$ and $-135^\circ$, respectively.}


Note that Eq.~(B.15) is meaningful only if the argument inside the arctangent is dimensionless. This condition is satisfied when normalized units, $\hat{R}$, $\hat{\varepsilon}$, $\hat{F_1}$, and $\hat{F_2}$ as described in Appendix A, are used where all these linear parameters have units of month$^{-1}$.


\appendix[C]
\appendixtitle{Analytical Expressions for the Power Spectral Density of the Niño index}

The simplest form of the linear RO, Eq.~(B1), including noise terms, can be expressed in the frequency domain over $-\infty$ and $\infty$ (two-sided spectrum) as: 

\begin{linenomath*}
\begin{equation}
i\omega \tilde{T}_E (\omega) =
R \tilde{T}_E (\omega) + F_1 \tilde{h} (\omega) + \sigma_T \tilde{\xi}_T (\omega)
\end{equation}

\begin{equation}
i\omega \tilde{h} (\omega) =
-\varepsilon \tilde{h} (\omega) - F_2 \tilde{T}_E (\omega) + \sigma_h \tilde{\xi}_h (\omega)
\end{equation}
\end{linenomath*}

\noindent Solving for $\tilde{T}_E (\omega)$, we get:

\begin{linenomath*}
\begin{equation}
\tilde{T}_E (\omega) =
\frac{ F_1 \sigma_h \tilde{\xi}_h (\omega) + (i\omega + \varepsilon)\sigma_T \tilde{\xi}_T (\omega) }
{ (i\omega - R)(i\omega + \varepsilon) + F_1 F_2 }
\end{equation}
\end{linenomath*}

\noindent Assuming $|\tilde{\xi}_h (\omega) \tilde{\xi}_T (\omega)| = 0$, the power spectral density is:

\begin{linenomath*}
\begin{equation}
|\tilde{T}_E (\omega)|^2 =
\frac{ F_1^2 \sigma_h^2 |\tilde{\xi}_h (\omega)|^2 
+ (\omega^2 + \varepsilon^2) \sigma_T^2 |\tilde{\xi}_T (\omega)|^2 }
{ (-\omega^2 - R\varepsilon + F_1 F_2)^2 + \omega^2 (\varepsilon - R)^2 }
\end{equation}
\end{linenomath*}

\noindent For white noise: if $\tilde{\xi}_T$ and $\tilde{\xi}_h$ are white noise with unit variance, 
$|\tilde{\xi}_T (\omega)|^2 = |\tilde{\xi}_h (\omega)|^2 = \frac{N_0}{2}=\text{constant}$. The factor $\frac{1}{2}$ accounts for the double counting of energy in the two-sided spectrum compared to the one-sided power spectrum (over 0 and $\infty$). The power spectral density becomes: 

\begin{linenomath*}
\begin{equation}
|\tilde{T}_E (\omega)|^2 = \frac{N_0}{2} \cdot
\frac{ F_1^2 \sigma_h^2 + (\omega^2 + \varepsilon^2) \sigma_T^2 }
{ (-\omega^2 - R\varepsilon + F_1 F_2)^2 + \omega^2 (\varepsilon - R)^2 }
\end{equation}
\end{linenomath*}

\noindent For red noise: If $\xi_T$ and $\xi_h$ are red noise, expressed as

\begin{linenomath*}
\begin{equation}
\frac{d\xi_T}{dt} = -m_T \xi_T + \sqrt{2m_T} \, w_T
\end{equation}

\begin{equation}
\frac{d\xi_h}{dt} = -m_h \xi_h + \sqrt{2m_h} \, w_h
\end{equation}
\end{linenomath*}

\noindent where the factors $\sqrt{2m_T}$ and $\sqrt{2m_h}$ are introduced so that the noise terms have unit variance under decorrelation times of $m_T$ and $m_h$, respectively. The power spectral densities for these noise terms are:

\begin{linenomath*}
\begin{equation}
|\tilde{\xi}_T (\omega)|^2 = \frac{2m_T}{\omega^2 + m_T^2}
\end{equation}

\begin{equation}
|\tilde{\xi}_h (\omega)|^2 = \frac{2m_h}{\omega^2 + m_h^2}
\end{equation}
\end{linenomath*}

\noindent For this case, the power spectral density of the Niño index becomes:

\begin{linenomath*}
\begin{equation}
|\tilde{T}_E (\omega)|^2 = \frac{N_0}{2} \cdot
\frac{ F_1^2 \sigma_h^2 \frac{2}{m_h\left(\frac{\omega^2}{m_h^2} + 1\right)} 
+ (\omega^2 + \varepsilon^2) \sigma_T^2 \frac{2}{m_T\left(\frac{\omega^2}{m_T^2} + 1\right)} }
{ (-\omega^2 - R\varepsilon + F_1 F_2)^2 + \omega^2 (\varepsilon - R)^2 }
\end{equation}
\end{linenomath*}

\bibliographystyle{ametsocV6}
\bibliography{ENSO}

\end{document}